\documentclass[12pt]{article}
\usepackage{amsfonts,amssymb,amsmath}
\usepackage{epic,eepic}
\usepackage{theorem,cite}
\newtheorem{definition}{Definition}[section]

\newtheorem{proposition}[definition]{Proposition}
\newtheorem{property}[definition]{Property}
\newtheorem{theorem}[definition]{Theorem}
\newtheorem{corollary}[definition]{Corollary}
\newtheorem{lemma}[definition]{Lemma}

\theorembodyfont{\rmfamily}

\numberwithin{equation}{section}

\def\cA{{\cal A}}          \def\cB{{\cal B}}          
          \def\cE{{\cal E}}          
          \def\cH{{\cal H}}          
                    \def\cL{{\cal L}}
\def\cM{{\cal M}}          \def\cN{{\cal N}}          
\def\cP{{\cal P}}                    \def\cR{{\cal R}}
\def\cS{{\cal S}}          \def\cT{{\cal T}}          \def\cU{{\cal U}}
                    
\def\cY{{\cal Y}}

\def\fa{{\mathfrak a}}
\def\fb{{\mathfrak b}}
\def\fg{{\mathfrak g}}
\def\fh{{\mathfrak h}}

\newcommand{\CC}{{\mathbb C}}

\newcommand{\II}{{\mathbb I}}

\newcommand{\ZZ}{{\mathbb Z}}
\newcommand{\eps}{{\varepsilon}}

\newcommand{\id}{\II^{\otimes\enne}}

\newcommand{\atopn}[2]{\genfrac{}{}{0pt}{}{#1}{#2}}

\newcommand{\finproof}{{\hfill \rule{5pt}{5pt}\\}}

\newcommand{\bs}[1]{{\boldsymbol{#1}}}
\newcommand{\wh}[1]{{\widehat{#1}}}
\newcommand{\wt}[1]{{\widetilde{#1}}}

\newcommand{\qmbox}[1]{{\qquad\mbox{#1}\quad}}
\def\qmbox#1{\qquad\mbox{#1}\quad}

\def\tr{\mathop{\rm tr}\nolimits}

\newcommand{\enne}{{\cal N}}
\newcommand{\emme}{{\cal M}}

\newcommand{\nonu}{\nonumber\\}

\begin{document}
\pagestyle{empty}

\markright{\today\dotfill Version multiplicative\dotfill }
%
\null
\vfill

\begin{center}

{\Large \textsf{Spectrum and Bethe ansatz equations
for the $U_{q}(gl(\enne))$\\[1.2ex]
closed and open spin chains
in any representation }}

\vspace{2.4em}

{\large D. Arnaudon$^a$,
  N.~Cramp{\'e}$^{b}$\footnote{
  nc501@york.ac.uk, doikou@lapp.in2p3.fr,
  frappat@lapp.in2p3.fr, ragoucy@lapp.in2p3.fr\label{foot:1}},
  A.~Doikou$^{a\ref{foot:1}}$, L. Frappat$^{ac\ref{foot:1}}$,
  {E}. Ragoucy$^{a\ref{foot:1}}$}

\vspace{1.2em}

\emph{$^a$ Laboratoire d'Annecy-le-Vieux de Physique Th{\'e}orique}

\emph{LAPTH, UMR 5108, CNRS and Universit{\'e} de Savoie}

\emph{B.P. 110, F-74941 Annecy-le-Vieux Cedex, France}

\vspace{2.1ex}

\emph{$^b$ University of York, Department of Mathematics}

\emph{Heslington, York YO10 5DD, United Kingdom}

\vspace{2.1ex}

\emph{$^c$ Member of Institut Universitaire de France}

\end{center}

\vfill

\begin{abstract}
We consider the $N$-site $U_{q}(gl({\cal N}))$ integrable spin chain with
periodic and open diagonal soliton-preserving boundary conditions. By
employing analytical Bethe ansatz techniques we are able to determine the
spectrum and the corresponding Bethe ansatz equations for the general case,
where each site of the spin chain is associated to any representation of
$U_{q}(gl({\cal N}))$.

In the case of open spin chain, we study finite dimensional
representations of the quantum reflection algebra, and prove in full
generality that the
pseudo-vacuum is a highest weight of the monodromy matrix.

For these two types of spin chain, we study the (generalized)
"algebraic" fusion procedures, which amount to construct the quantum
contraction and the Sklyanin determinant for the $U_{q}(\wh{gl}({\cal N}))$
and quantum reflection
algebras. We also determine the symmetry algebra of these two 
types of spin chains, including general $K$ and $K^+$ diagonal matrices 
for the open case.

The case of open spin chains with soliton non-preserving boundary conditions
is also presented in the framework of quantum twisted Yangians.
The symmetry algebra of this spin chains is studied.
We also give an exhaustive classification of the invertible matricial 
solutions to the corresponding reflection equation.

\end{abstract}

\vfill
\vfill
\centerline{\textit{Dedicated to our friend D. Arnaudon}}
\vfill

\null\quad MSC: 81R50, 17B37 \hfill
PACS: 02.20.Uw, 03.65.Fd, 75.10.Pq\qquad\\

\texttt{math-ph/0512037}\hfill {LAPTH-1133/05}\\
\rightline{December 2005}

\baselineskip=16pt

\newpage

\pagestyle{plain} \setcounter{page}{1}

\setcounter{tocdepth}{2}
\tableofcontents
\newpage

\section{Introduction}

For a long time, integrable systems, in particular spin chains
models, have attracted much attention. The
importance of these models relies on the fact that nonperturbative
expressions of physical values (eigenstates, correlation
functions,...) may be obtained exactly. Due to this property,
numerous applications have been obtained in
different domains of physics (condensed matter, string
theory,...) as well as mathematics (quantum groups,...).
Among the different
approaches used to solve integrable problems, the Bethe ansatz
\cite{bet} has been historically introduced to obtain eigenstates of the
XXX model proposed by Heisenberg \cite{heisen}. Then, various
generalisations of this ansatz have been successfully constructed and 
applied.
In this paper, we
shall use a generalisation of this method, called analytical Bethe
ansatz \cite{reshe}, to find the spectrum of the periodic
$\cU_q(gl(\enne))$ spin chain (XXZ model) where at each site the
$\cU_q(gl(\enne))$ representation may be different. 
The construction of these models
follows the same lines that the one done previously for the
$gl(\enne)$ spin chain \cite{byebye} and is based on the finite
irreducible representations of $\cU_q(\wh{gl}(\enne))$. Then, to find
the energy spectrum, we need to know a particular eigenvector (which
is simply the highest weight of the chosen representation), to determine
the symmetry and to obtain the explicit form of the
$\cU_q(\wh{gl}(\enne))$ center. The knowledge of these algebraic data is
sufficient to compute the full spectrum of the so-called transfer
matrix (which gathers $N$ Hamiltonians for the $N$ sites spin chain).

More recently, the introduction of boundaries which preserve the
integrability of well-known models has been also investigated
\cite{cherednik,sklyanin}. In the present context, the construction of the
$\cU_q(gl(\enne))$ spin chain with non-periodic boundaries consists in
studying the representations of $\cU_q(\wh{gl}(\enne))$ subalgebras 
instead of the whole algebra. In particular, we study the quantum
reflection algebra and the quantum twisted Yangian which are
respectively associated to the soliton preserving (SP) and soliton
non-preserving (SNP) boundaries. In the SP case, we determine the
symmetry algebra and the spectrum of the corresponding transfer
matrix, using the analytical Bethe ansatz. This case is very similar
to the closed spin chains case. In the SNP case, we present the 
algebraic setup and classify the matricial solutions to the
corresponding reflection equation. However, the
absence of diagonal solution prevents us from completing the study, 
for one cannot find a reference state 
(pseudo-vacuum).

The plan of the paper is as follows. In section \ref{sect:alg}, we
introduce the algebraic structures which will be needed for the
study of spin chains models. They consist in the quantum affine
algebra (section \ref{sect:Uq}) and the quantum reflection algebra
(section \ref{sect:qrefl}), both of them based on the $gl(\enne)$
algebra. We will also remind the irreducible finite-dimensional
representations of the quantum affine algebra, and study those of
the quantum reflection algebra. To our knowledge, this latter study
is new. Then, in section \ref{sect:spch}, we construct the spin
chains associated to these algebraic structures: closed spin chains
for quantum affine algebra (section \ref{sect:closed}) and 
SP open spin chain for the quantum reflection algebra
(section \ref{sect:open}).  We compute their spectrum, through
the use of the analytical Bethe ansatz. Finally, we introduce
in section \label{qYang-SNP} the framework needed for the study of
SNP open spin chains. The algebraic setting
consists in the quantum twisted Yangian (section \ref{sect:qYtw}).
SNP open spin chains are introduced in
section \ref{sect:SNP}. We classify the matrices obeying the
corresponding reflection equation. However, due to the absence of
diagonal solution, the usual Bethe ansatz does not work: we argue on
the complete treatment of these spin chains. Appendices are devoted
to some properties of the $R$ matrices involved in the study
(appendix \ref{sect:Rmatrix}), as well as to the fusion procedures
used for the analytical Bethe ansatz. At the algebraic level, these
fusion procedures amount to construct the quantum contraction and
the Sklyanin determinant for the $\cU_{q}(\wh{gl}({\cal N}))$ and
quantum reflection algebras (appendices \ref{sect:fQ} and
\ref{sect:gefu}).

\section{Algebraic structures\label{sect:alg}}
In this section, we describe the  algebraic framework
needed for the construction of the different spin chains which will be
presented in the next section. Depending on the boundary conditions
we will impose on the spin chain, two types of algebras will show
up: the quantum affine algebra and the quantum reflection algebra.
They both rely on the $R$ matrix of $\cU_{q}(\wh{gl}({\cal
N}))$, and we have gathered the different properties of this matrix
in appendix \ref{sect:Rmatrix}.

\subsection{The quantum affine algebra $\cU_{q}(\wh{gl}({\cal N}))$\label{sect:Uq}}
\subsubsection{Definitions \label{sect:defClsd}}

One starts from the exchange relations of finite dimensional $\cU_{q}({gl}({\cal N}))$
algebra:
\begin{eqnarray}
  R_{12}\,{\cal L}^{+}_{1}{\cal L}^{+}_{ 2} = {\cal L}^{+}_{2} {\cal
  L}^{+}_{1}\,R_{12}\quad ;\quad R_{12}\,{\cal L}^{+}_{1}{\cal L}^{-}_{ 2}
  = {\cal L}^{-}_{2}{\cal L}^{+}_{1}\, R_{12}\quad ;\quad R_{12}\,{\cal
  L}^{-}_{1}{\cal L}^{-}_{ 2} = {\cal L}^{-}_{2}{\cal L}^{-}_{1}\,R_{12}
\end{eqnarray}
where $R_{12}$ is given by (\ref{Ruqfin}) and ${\cal L}^{+}$, ${\cal
L}^{-}$ are upper (lower) triangular matrices defining $\cU_{q}({gl}({\cal
N}))$, i.e.
\begin{equation}
  {\cal L}^{+} = \sum_{1\leq a\leq b \leq{\cal N}} E_{ab} \otimes \ell_{ab}^+,
  ~~~~~{\cal L}^{-} = \sum_{1\leq b\leq a \leq{\cal N}} E_{ab} \otimes
  \ell_{ab}^{-}
\quad\mbox{with}\quad \ell_{ab}^+,~\ell_{ab}^{-} \in \cU_{q}(gl({\cal
N}))\,.   \label{lh}
\end{equation}
There are supplementary relations between the diagonal elements, namely
\begin{equation}
\ell_{aa}^+\ell_{aa}^-=\ell_{aa}^-\ell_{aa}^+=1\,.
\end{equation}

The algebra $\cU_{q}(\wh{gl}({\cal N}))$ (noted for brevity ${\wh\cU_{q}}$)
is defined by the following fundamental relations, known as FRT
relations \cite{FRT,difre}
\begin{eqnarray}
  R_{12}(\frac{z}{w})\ {\cal L}^\pm_{1}(z)\ {\cal L}^\pm_{ 2}(w) 
  &=& {\cal L}^\pm_{2}(w)\ {\cal L}^\pm_{1}(z)\ R_{12}(\frac{z}{w}),
  \label{rtt}\\
  R_{12}(\frac{z}{w}\,q^c)\ {\cal  L}^+_{1}(z)\ {\cal L}^-_{ 2}(w) 
  &=& {\cal L}^-_{2}(w)\ {\cal L}^+_{1}(z)\ R_{12}(\frac{z}{w}\,q^{-c}),
  \label{rtt2}
 \end{eqnarray}
where, as usual in auxiliary spaces formalism,
$\cL^\pm_{1}(z)=\cL^\pm(z)\otimes\II_{\enne}\in
End(\CC^\enne)\otimes End(\CC^\enne)\otimes {\wh\cU_{q}}$,
$\cL^\pm_{2}(z)=\II_{\enne}\otimes \cL^\pm(z)\in
End(\CC^\enne)\otimes End(\CC^\enne)\otimes {\wh\cU_{q}}$ and
$R_{12}(\frac{z}{w})\in
End(\CC^\enne)\otimes End(\CC^\enne)$ is given by (\ref{r}).
The space $End(\CC^\enne)$ is known as the auxiliary space and $c$ is
the central charge (which will be set to zero below).\\
${\cal L}^\pm(z)\in End(\CC^\enne)\otimes {\wh\cU_{q}}$ gathers the generators
$L_{ab}^{(\pm n)}$ as
\begin{equation}
{\cal L}^\pm(z)=\sum_{n=0}^{+\infty}\sum_{a,b=1}^\enne z^{\pm
2n}\,E_{ab}\otimes  L_{ab}^{(\pm n)}=\sum_{a,b=1}^\enne
E_{ab}\otimes  L^\pm_{ab}(z),
\label{serieL}
\end{equation}
with the constraints
\begin{eqnarray}
    &&L_{ab}^{(+ 0)}=L_{ba}^{(- 0)}=0\,,\qquad 1\leq a<b\leq\enne
    \label{eq:L-triang}\\
    &&L_{aa}^{(+0)}\,L_{aa}^{(-0)}=L_{aa}^{(-0)}\,L_{aa}^{(+0)}=1
    \,,\qquad a=1,\ldots,\enne. \label{eq:L-diag}
\end{eqnarray}
Note that, due to the convention we take for the $R$ matrix, the
$\cL^\pm(z)$ operators are even in $z$, so that the series expansion
involves even integers only.

Remark also that the "level zero" generators $L_{ab}^{(+ 0)}$ and
$L_{ab}^{(- 0)}$ span a finite dimensional algebra which is nothing
but the $\cU_{q}({gl}({\cal N}))$ algebra: in the following, we will
identify $L_{ab}^{(\pm 0)}$ with $\ell^\mp_{ab}$.

${\wh\cU_{q}}$ is endowed with a coproduct $\Delta: {\wh\cU_{q}} \to {\wh\cU_{q}}
\otimes {\wh\cU_{q}}$
\begin{eqnarray}
&& \Delta({\cal L}^\pm(z))=
\cL^\pm_{01}(zq^{\pm\frac{c_{2}}{2}})\,
\cL^\pm_{02}(zq^{\mp\frac{c_{1}}{2}}),\\
&& \Delta(c)=c_1+c_2
\end{eqnarray}
where $0$ denotes the auxiliary space, and $1,2$ label the copies of
${\wh\cU_{q}}$ in which act the operators. For instance,
$\cL_{02}^\pm(z)$ acts trivially in the first copy of ${\wh\cU_{q}}$,
while $c_{1}=c\otimes 1$ and $c_{2}=1\otimes c$.

When $c=0$ (which will be always the case in section
\ref{sect:spch}), the coproduct reduces to
\begin{equation}
  \Delta(L^\pm_{ab}(z))= \sum_{c=1}^{{\cal N}} L^\pm_{ac}(z)
  \otimes L^\pm_{cb}(z), ~~~a,\ b \in \{ 1,
  \ldots, {\cal N} \}.
  \label{cob1}
\end{equation}
More generally, one defines recursively for $N\geq 2$
\begin{equation}
\Delta^{(N+1)}=(id^{\otimes (N-1)}\otimes \Delta)\circ\Delta^{(N)}\
:\ {\wh\cU_{q}} \to {\wh\cU_{q}}^{\otimes (N+1)}
\end{equation}
with $\Delta^{(2)}=\Delta$ and $\Delta^{(1)}=id$. The map $\Delta^{(N)}$
 is also a morphism, i.e.
 \begin{equation}
  \cT^\pm_{0}(z) = \Delta^{(N)}\left({\cal L}^\pm(z)\right)=
  {\cal L}^\pm_{01}(z)\ {\cal L}^\pm_{02}(z) \ldots {\cal
  L}^\pm_{0\ N-1}(z)\ {\cal L}^\pm_{0N}(z) \label{mono}
\end{equation}
 also obey the relations (\ref{rtt})-(\ref{rtt2}) and the constraints
 (\ref{eq:L-triang})-(\ref{eq:L-diag}).
As usual, the indices $i\in \{1, \ldots, N \}$, here labeling
copies of ${\wh\cU_{q}}$,  are suppressed from $\cT$, and we only keep the
index $0$ corresponding to the `auxiliary space'.

As we shall see below, in the context of spin chains, $\cT^\pm(z)$ will be seen (after being
represented) as monodromy
matrices. For the moment, we remark that this
construction is valid at the algebra level, i.e. before any choice of
representations, a property which justifies the name of universal (or
algebraic)
monodromy matrices for $\cT^\pm(z)$. Correspondingly, we introduce
two universal transfer matrices
\begin{equation}
  t^\pm(z) =tr_{0}\ \cT^\pm_{0}(z) \label{transfer}
\end{equation}
which are elements of ${\wh\cU_{q}}^{\otimes N}$. It can be then
shown via (\ref{rtt})-(\ref{rtt2}) that \cite{faddeev}
\begin{equation}
  \Big [ t^\eps(z),\ t^{\eps'}(w)\Big] =0\,,\qquad
  \eps,\eps'\in\{+,-\}. \label{com}
\end{equation}
We remark that we have \textit{two} transfer matrices: in fact, we
will see below that the construction of a spin chain needs only
$\cT^+(z)$ (hence only $t^+(z)$), because the
construction  based on $\cT^-(z)$ (and
$t^-(z)$) is equivalent.

\subsubsection{Symmetry of the transfer matrix}

The relations (\ref{rtt})-(\ref{rtt2}), using the form
(\ref{r-joli}) of the $R$ matrix, can be rewritten as (when $c=0$):
\begin{eqnarray}
{[\cT_{1}(z)\,,\,\cT_{2}(w)]} &=& \Big\{
\cT_{2}^q(z)\cT_{1}^{\bar q}(w)-\cT_{2}(w)\cT_{1}(z) \Big\}P_{12}^q
+\frac{\fa(\frac{z}{w})}{\fb(\frac{z}{w})}\,\Big\{ \cT_{2}(w)\cT_{1}(z)
-\cT_{2}(z)\cT_{1}(w)\Big\}\cP \qquad\quad
\end{eqnarray}
where $\cT(x)$ stands for $\cT^{+}(x)$ or $\cT^-(x)$, $\cT^q(z)$ for
its $q$-deformed
operator defined
in relation (\ref{def:q-op}) and $\bar q=q^{-1}$.

Taking the trace in the auxiliary space 1 leads to
\begin{equation}
{[t(z)\,,\, \cT(w)]} = [\cT^{ q}(z)
-\frac{\fa(\frac{z}{w})}{\fb(\frac{z}{w})}\,\cT(z)\,,\,\cT(w)]\,.
\end{equation}
Then, using the expansions
\begin{equation}
\cT^\pm(w)=\cL^\mp+o(w^{\pm2})\qmbox{,}
\frac{\fa(\frac{z}{w})}{\fb(\frac{z}{w})}=q^{\pm1}+o(w^{\pm2})
\end{equation}
 one can compute the action of the $\cU_{q}({gl}({\cal N}))$ generators
 on the transfer matrix:
\begin{equation}
{[t(z)\,,\, \cL^\pm]} = [\cT^{ q}(z)-q^{\mp1}\cT(z)\,,\,\cL^{\pm}]\,.
\end{equation}
In particular, considering the diagonal terms, using the fact that
$\cL^\pm$ is triangular and the commutation relations
$[\ell^\pm_{ii}\,,\,\cT_{jj}(z)]=0$, one gets the following result:
\begin{property}
All the Cartan generators of
the finite Lie algebra $\cU_{q}({gl}({\cal N}))$ commutes with the
universal transfer matrix,
\begin{equation}
{[\ell^\pm_{ii}\,,\, t(z)]} =0\,,\ i=1,\ldots,\enne\,.
\end{equation}
Thus, they generate a $U(1)^\enne$ symmetry
algebra for the closed spin chains.
\end{property}

\subsubsection{Representations and evaluation map\label{Uqrep}}

Since the spin chain interpretation will be possible through the
use of representations of the algebra we consider, we now describe them.
The irreducible finite dimensional representations of $\cU_{q}(\wh{gl}({\cal N}))$
are built from those of $\cU_{q}({gl}({\cal N}))$ using the evaluation
morphism.

It is easy to show  that the (evaluation) map
\begin{eqnarray}
  && {\cal L}^+(z) \mapsto {\cal L}^{-} - z^2
  {\cal L}^{+} \label{def-juste:L+z}\\
  && {\cal L}^-(z) \mapsto {\cal L}^{+} - z^{-2}
  {\cal L}^{-}  \label{def-juste:L-z}
\end{eqnarray}
defines a homomorphism from $\cU_{q}(\wh{gl}({\cal N}))$ (at $c=0$)
to $\cU_{q}({gl}({\cal N}))$.
Note that $\cL^+(z)$ and $\cL^-(z)$ are represented in the
same way (up to a multiplicative function of $z$). In fact, when $c=0$
(which is always the case for finite dimensional
representations), they play the same role and we will rather work with
\begin{equation}
    {\cal L}(z)=  z  {\cal L}^{+} - z^{-1}{\cal L}^{-}\,.
  \label{lh0}
\end{equation}
Let us remark however that both for the mathematical framework, or for
the analyticity properties used in Bethe ansatz, one should work with 
$\cL^{+}(z)$ rather than $\cL(z)$. Unfortunately, it is this latter
notation which is used in spin chains context, so that we stick to
it. We will come back to this point in section \ref{sect:qrefl}.

The construction of finite dimensional irreducible representations of
$\cU_{q}(\wh{gl}({\cal N}))$ is rooted in this evaluation morphism. Thanks to
this map, each representation of $\cU_{q}({gl}({\cal N}))$ can be lifted to a
representation of $\cU_{q}(\wh{gl}({\cal N}))$, and indeed, it can be shown
that all (up to twist by sign automorphisms) finite dimensional irreducible
representations of
$\cU_{q}(\wh{gl}({\cal N}))$ can be constructed in this way (when $q$
is neither a root of unity, nor zero),
see for instance \cite{bookChPr}, section 12.2.B.
Since we will
use this property in the next section to build the different spin chains
based on $\cU_{q}(\wh{gl}({\cal N}))$, we describe this construction.

We start with a highest weight finite-dimensional irreducible
representation of $\cU_{q}(gl(\enne))$, $M({\bs \varpi})$, with highest
weight ${\bs \varpi} = (\varpi_1,\dots,\varpi_\enne)$ and associated to the
highest weight vector $v$. This highest weight vector obeys
\begin{eqnarray}
  && \ell_{ab}^{-} v = 0 \;, \qquad 1 \le b<a \le \enne \\
  && \ell_{aa}^{-} v = \eta_{a}\,q^{-\varpi_{a}} v \;, \qquad 1 \le
  a \le \enne
\end{eqnarray}
where (when $q$ is generic) $\varpi_1,\dots,\varpi_\enne$ are real
numbers
with $\varpi_{a}-
\varpi_{a+1}\in\ZZ_{+}$ and $\eta_{a}=\pm1,\pm i$ (see \cite{rosso} and
theorem \ref{theo:repUq} below).
When $q$ is a root of unity, the parameters
$\varpi_1,\dots,\varpi_\enne$ obey more general relations. In what
follows, we will formally redefine $\varpi_{a}$ in such  way that
$\eta_{a}=+$, and assume that the weights $\varpi_{a}$ obey the
conditions for the representation to be irreducible, finite
dimensional and highest weight (including the cases where $q$ is a root of unity).
Then, the repetitive action of $\ell_{ab}^{+}$, $1\leq a<b\leq \enne$,
generates the other states of the representation.
\\
Using the evaluation homomorphism (\ref{lh0}), one infers from $M({\bs \varpi})$
an
irreducible finite dimensional highest weight
representation $M_{z}({\bs \varpi})$ for $\cU_{q}(\wh{gl}(\enne))$:
\begin{eqnarray}
  && L_{ab}(z) v = 0 \;, \qquad 1 \le b < a \le \enne \\
  && L_{aa}(z) v = \left( zq^{\varpi_{a}} -
  z^{-1}q^{-\varpi_{a}} \right) v \;, \qquad 1 \le a \le \enne
\end{eqnarray}
More generally, using the morphism $\Delta^{(N)}$, one constructs the tensor
product of such representations $M_{zq^{a_{1}}}({\bs \varpi}^{(1)})\otimes
\ldots\otimes M_{zq^{a_{N}}}({\bs
\varpi}^{(N)})$, with so-called `inhomogeneity' parameters $a_{n}$
and highest weights ${\bs \varpi}^{(n)}=(\varpi^{(n)}_{1},\ldots,\varpi^{(n)}_{\enne})$,
$n=1,\ldots,N$.
This tensor product infers a
representation for the universal monodromy matrices
$\cT^\pm(z)=\Delta^{(N)}{\cal L}^\pm(z)$:
\begin{eqnarray}
&&T_{ab}(z)\,\omega = 0 \;,\qquad 1 \le b < a \le \enne \\
&&  T_{aa}(z)\,\omega = \prod_{n=1}^{N} \left(z
  q^{a_{n}+\varpi_{a}^{(n)}} - z^{-1}
  q^{-a_{n}-\varpi_{a}^{(n)}} \right)\, \omega
  \equiv P_{a}(z)\, \omega
  \label{drinP}
\end{eqnarray}
where $\omega=\otimes_{i=1}^{N}v^{(i)}$, $v^{(i)}$ is the highest weight
vector of $M({\bs \varpi}^{(i)})$, $i=1,\ldots,N$. The functions
$P_{a}(z)$ are related to the Drinfel'd polynomials $D_{a}(z)$ characterizing the
representation in the usual way, namely
\begin{equation}
    \frac{P_{a}(z)}{P_{a+1}(z)} = \frac{D_{a}(q\,z)}{D_{a}(z)}.
    \label{polyDrin}
\end{equation}
Being a highest weight vector of the monodromy
matrix, $\omega$ is an eigenvector of the transfer matrix:
\begin{equation}
t(z)\,\omega=\Lambda^{0}(z)\,\omega \qmbox{with}
\Lambda^{0}(z) = \sum_{a=1}^\enne P_{a}(z)
\label{lambda0}
\end{equation}

In a mathematical context, the quantum contraction or quantum
determinant generate central elements of the algebra
$\cU_{q}(\wh{gl}({\cal N}))$. In the spin chain context, they allow
one to obtain constraints on the eigenvalues of the problem. The
different useful formulae are gathered in appendices \ref{sect:fQ}
and \ref{sect:gefu}. For a given representation, the quantum
contraction and the quantum determinant of $\cT(z)$ take
respectively the simple
forms
\begin{equation}
\label{eq:qdetrep} \delta(\cT(z))=1
\qmbox{and}qdet\,\cT(z)=\prod_{a=1}^\enne P_a(zq^{\enne-a})\,.
\end{equation}
To prove these formulae, we have used relations (\ref{eq:qcontL})
and (\ref{qdet2}) applied on the highest weight vector $\omega$, the
statement that the quantum contraction and quantum determinant are
proportional to the identity matrix (because it belongs to the
center of $\cU_{q}(\wh{gl}({\cal N}))$) and the convention that the
length of the identity permutation is zero.

\subsection{The quantum reflection algebra \label{sect:qrefl}}

\subsubsection{Definitions\label{sect:defOpen}}

The quantum reflection algebra $\cR$ is constructed as a coideal
subalgebra of $\cU_{q}(\wh{gl}({\cal N}))$. This general
construction can be done for "any" quantum group defined by an
FRT relation of  type (\ref{rtt}) provided the $R$ matrix obeys
 relation (\ref{supR}).
In a physical context, this construction reflects the possibility,
starting from a periodic (closed) spin chain, to build a new chain
with an (integrable) boundary (open spin chain). We focus in this
subsection on (some of) the mathematical aspects of the
construction, postponing the physical contents to the next section.

As a first step, one defines
\begin{equation}
    \cB(z)={\cal L}^+(zq^{-\frac{c}{2}})
    \,{\cal L}^-(z^{-1}q^{\frac{c}{2}})^{-1},
\label{defB}
\end{equation}
where ${\cal L}^\pm(z)$ obey the relations
(\ref{rtt})-(\ref{rtt2}) and the existence of ${\cal L}^-(z^{-1})^{-1}$,
understood as a series expansion, is guaranteed by relations
(\ref{eq:L-diag}). Note that the forms (\ref{def-juste:L+z}) and
(\ref{def-juste:L-z}) ensure that $\cB(z)$ is analytical. This point
has to be related to the remark made in section \ref{Uqrep} on analyticity
conditions. However, for $c=0$, one
can loosely rewrite $\cB(z)$ as $\cL(z)\,\cL(z^{-1})^{-1}$, with $\cL(z)$
defined by (\ref{lh0}).

It is not difficult to show that (the
generating function)
$\cB(z)$ defines a subalgebra $\cal R$, called the (quantum)
reflection algebra, whose exchange relations take the form
\begin{equation}
  R_{12}(\frac{z}{w})\,\cB_{1}(z)\
  R_{21}(zw)\,\cB_{2}(w)= \cB_{2}(w)\
  R_{12}(zw)\,\cB_{1}(z)\
  R_{21}(\frac{z}{w}).
  \label{re}
\end{equation}
To prove (\ref{re}) starting from (\ref{defB}),
one needs the supplementary constraint on $R$:
\begin{equation}
    R_{12}(z)R_{21}(w)=R_{12}(w)R_{21}(z),
    \label{supR}
\end{equation}
which is indeed satisfied by the $R$ matrix of $\cU_{q}(\wh{gl}({\cal N}))$.

Due to the series expansions (\ref{serieL}), $\cB(z)$ can be expanded as
\begin{equation}
\cB(z)=\sum_{n=0}^\infty\sum_{a,b=1}^\enne z^{2n}\, E_{ab}\otimes
B_{ab}^{(n)} = \sum_{a,b=1}^\enne  E_{ab}\otimes
B_{ab}(z) = \sum_{n=0}^\infty  z^{2n}
\cB^{(n)}
\label{serieB}
\end{equation}
The $\cR$ subalgebra is  a left coideal (see e.g. \cite{ZFbound}) of the
starting algebra ${\wh\cU_{q}}=\cU_{q}(\wh{gl}({\cal N}))$:
\begin{equation}
  \Delta({\cal B}(z)) =  {\cal
  L}^{+}_{01}(zq^{-\frac{c_{1}}{2}})\,
  {\cB}_{02}(zq^{-c_{1}})\, {\cal
  L}^{-}_{01}(q^{\frac{c_{1}}{2}}z^{-1}) ^{-1}\ \in\ {\wh\cU_{q}}\otimes\cR
\label{coc}
\end{equation}
where we have used the index $0$ for the auxiliary space, and $1,2$
for the two copies of ${\wh\cU_{q}}$. This formula
 is easily generalized for $\Delta^{(N)}$:
\begin{eqnarray}
\Delta^{(N)}\cB(z) =
\cL^{+}_{01}(zq^{-\frac{c_{1}}{2}})\cL^{+}_{02}(zq^{-c_{1}-\frac{c_{2}}{2}})\ldots
\cL^{+}_{0N}(zq^{-c_{1}-c_{2}-\ldots-c_{N-1}-\frac{c_{N}}{2}})\nonumber\\
\qquad\qquad\times
\cL^{-}_{0N}(q^{c_{1}+c_{2}+\ldots+c_{N-1}+\frac{c_{N}}{2}}z^{-1})^{-1}\ldots
\cL^{-}_{02}(q^{c_{1}+\frac{c_{2}}{2}}z^{-1})^{-1}
\cL^{-}_{01}(q^{\frac{c_{1}}{2}}z^{-1})^{-1}\,.
\label{KN}
\end{eqnarray}

A more general solution of (\ref{re}) is then given by \cite{sklyanin}:
\begin{equation}
  {\cal B}(z) = {\cal L}^+(zq^{-\frac{c}{2}})\ K(z)\ {\cal
  L}^{-}(q^\frac{c}{2}z^{-1})^{-1},
  \label{gensol}
\end{equation}
where $K(z)$ is a $\CC$-valued matricial solution of the reflection
equation,
\begin{equation}
  R_{12}(\frac{z}{w})\ K_{1}(z)\
  R_{21}(zw)\ K_{2}(w)= K_{2}(w)\
  R_{12}(zw)\ K_{1}(z)\
  R_{21}(\frac{z}{w}). \label{reK}
\end{equation}
The relation (\ref{supR}) ensures that $K(z)=\II$ is a solution
to this equation.

Although  not always relevant for the mathematical structure of
the subalgebra, the $K$ matrix is pertinent at the physical level, and
encodes the boundary which is present in the open spin chain.
Indeed, from the mathematical point of view, one can replace
$K(z)$ by $U(z)K(z)U(\frac{1}{z})^{-1}$, where $U(z)$ is
any matricial solution of the relation 
$R_12(\frac{z}{w})\,U_1(z)\,U_2(w)=U_2(w)\,U_1(z)\,R_12(\frac{z}{w})$.
It amounts to a change $\cB(z)\to
U(z)\wt\cB(z)U(\frac{1}{z})^{-1}$, where $\wt\cB(z)$ is built on
$U(z)\cL(z)U(z)^{-1}$ instead to $\cL(z)$. All these changes being algebra 
automorphisms, the mathematical structure remains unaltered while the $K$
matrix (i.e. the boundary) has changed.

The diagonal solutions have been classified in \cite{dvgr2}, they take
the form
\begin{eqnarray}
\label{eq:Kdiag}
  && K(z; \xi) = \mbox{diag}\Big (\underbrace{\alpha, \ldots ,
  \alpha}_{\mbox{${\cal M}$}},\ \underbrace{ \beta, \ldots ,
  \beta}_{\mbox{${\cal N}-{\cal M}$}} \Big ),
  \\
  &&\alpha(z;\ \xi ) = z^2(z^2-\xi^2),
  \qquad \beta(z;\ \xi ) = 1-\xi^2 z^2,
\end{eqnarray}
where $\xi$ is a free parameter characterizing the boundary, and the
normalisation has been chosen to fulfill the series expansion
requirement.

The modes $\cB=\cB(0)$ generate a finite version of the reflection
algebra, whose exchange relations read
\begin{equation}
  R_{12}\,\cB_{1}\, R_{12}^{-1}\,\cB_{2}= \cB_{2}\,
  R_{21}^{-1}\,\cB_{1}\, R_{21}.
\end{equation}
Remark that, due to the presence of $K(z)$, the "zero mode"
$\cB$ does not generate a full $\cU_{q}(gl(\enne))$
algebra: we will come back on this point in section \ref{sect:finiteSalg}.

Note that one could also consider
\begin{equation}
\cB^\pm(z) = \cL^\pm(z)K(z)\cL^\pm(z^{-1})^{-1}\,,
\label{autreB}
\end{equation}
whose
commutation relations are also of the form (\ref{re}), but do not lead
to analytical entries for $\cB^\pm(z)$.
However, in the spin chain context,  we will be interested only with finite dimensional
representations, which have $c=0$, so that we will not distinguish the
three types of embeddings (we remind that $\cL^+(z)$ and
$\cL^-(z)$ are represented in the same way in evaluation
representations, see section \ref{Uqrep}). {From} now on, we will set $c=0$
and drop
the $\pm$ superscript.
Nevertheless, we want to
stress that only definitions of the form (\ref{defB}) lead to a coideal
subalgebra when $c\neq0$.

The (universal) monodromy matrix related to this algebra will
be defined using the morphism $\Delta^{(N)}$:
\begin{eqnarray}
\cB^N(z)\equiv\Delta^{(N)}\cB(z) &=&
\cL_{01}(z)\cL_{02}(z)\ldots\cL_{0N}(z)
K_{0}(z) \cL_{0N}(z^{-1})^{-1}\ldots
\cL_{02}(z^{-1})^{-1}\cL_{01}(z^{-1})^{-1}\nonu
&=& \cT_{0}(z)\,K_{0}(z)\,\cT_{0}(z^{-1})^{-1}\,.
\label{KgenN}
\end{eqnarray}
The monodromy matrix $\Delta^{(N)}\cB(z)$ also obeys  the
relation (\ref{re}). Upon representation, this monodromy matrix will
correspond to an open
spin chain with $N$ sites.\\
Finally we introduce the (universal) transfer matrix of the open spin chain
\cite{sklyanin}:
\begin{equation}
  b(z) = tr_{0}\ \Big( K_{0}^{+}(z)\,\cB^{N}(z)\Big),
  \label{transferSP}
\end{equation}
where
$K^{+}(z) $ is a solution of the dual reflection equation:
\begin{eqnarray}
  && R_{12}(\frac{w}{z})\ \Big(K_{1}^{+}(z)\Big)^{t_1}\ M_{1}^{-1}\
  R_{21}(\rho^{-2}z^{-1}w^{-1})\ M_{1}\ \Big(K_{2}^{+}(w)\Big)^{t_2}=
  \nonumber \\
  && \Big(K^{+}_{2}(w)\Big)^{t_2}\ M_{1}\
  R_{12}(\rho^{-2}z^{-1}w^{-1})\ M_{1}^{-1}\
  \Big(K_{1}^{+}(z)\Big)^{t_1}\ R_{21}(\frac{w}{z}).
  \label{red}
\end{eqnarray}
We have introduced $\rho
=q^\frac{\enne}{2}$ and the matrix $M$, defined in (\ref{def:M}).

Note that $M$ is a solution of (\ref{red}), which ensures that this
construction can be performed. In fact, from any solution $K(z)$ to
the reflection equation (\ref{re}), one can generate a solution
$K^{+}(z)$ to the dual equation (\ref{red}) through the combination
$K^{+}(z)=f(z)\Big(K((\rho z)^{-1})\Big)^t M$ where $f(z)$ is an
arbitrary function. Hence, starting from a diagonal solution for
$K(z)$, one can deduce a diagonal solution for $K^{+}(z)$:
\begin{eqnarray}
&&\hspace{-1cm}K^{+}(z) = q^{\enne+1}\,\mbox{diag}\Big
(\underbrace{q^{-2}\,\wt\alpha, q^{-4}\,\wt\alpha, \ldots ,
 q^{-2\cM_{+}}\,\wt\alpha}_{\mbox{${\cal M}_{+}$}},\ \underbrace{
 q^{-2\cM_{+}-2}\,\wt\beta, \ldots ,
  q^{-2\cN}\,\wt\beta}_{\mbox{${\cal N}-{\cal M}_{+}$}} \Big ),
    \qquad
  \label{diagKp}\\
\mbox{where}&&\wt\alpha(z;\ \xi_{+} ) = 1-(\rho\xi_{+}z)^2,
  \qquad \wt\beta(z;\ \xi_{+} ) = (\rho z)^{2}((\rho
  z)^{2}-\xi_{+}^2)\;.
\end{eqnarray}
The normalisation has been chosen in such a way that $K^{+}(z)$ has
analytical entries. Let us stress that, when considering a couple of
diagonal solutions given by (\ref{eq:Kdiag}) and (\ref{diagKp}), the
parameters $\xi$, $\cM$ and $\xi_{+}$, $\cM_{+}$ are not necessarily
related.

Again, it can be proved using only the
reflection equations (\ref{re}), (\ref{red}), and the properties of
the $R$ matrix, that \cite{sklyanin}
\begin{equation}
  \Big [ b(z),\ b(w) \Big ]=0\;.
\end{equation}
It will ensure that the open spin chain derived from (\ref{transferSP}) is also
integrable.

\subsubsection{Finite dimensional subalgebras of the reflection
algebra\label{sect:finiteSalg}}

{From} the reflection algebra, one deduces (in the limit
$w\to0$)\footnote{Note that this limit is well-defined only for the
form (\ref{gensol}), not for the alternative expressions (\ref{autreB}).}
\begin{eqnarray}
R_{12}\,\cB_{1}(z)\,R_{12}^{-1}\,{\cB}_{2}(0) =
{\cB}_{2}(0)\,R_{21}^{-1}\,\cB_{1}(z)\,R_{21}
\label{CR:B0}
\end{eqnarray}
where the above $R$ matrices are the finite ones of appendix \ref{sect:Rfini}.
It proves that $\cB(0)$ generates a subalgebra and that $\cB(z)$ is
one of its representation. To identify this subalgebra, one needs the
expression
\begin{equation}
K(0)=diag(\underbrace{0,\ldots,0}_{\emme},
\underbrace{1,\ldots,1}_{\enne-\emme})\,.
\end{equation}
This expression shows that $\cB(0)$
is constructed on the generators $\ell^{+}_{ij}$ and $\ell^{-}_{ij}$
with $i,j>\emme$.
It corresponds to the (finite dimensional version) of the reflection
algebra based on $\cU_{q}(gl(\enne-\emme))$.
In fact, this algebra is known to
be isomorphic to the $\cU_{q}(gl(\enne-\emme))$ algebra itself. This
is mainly due to the triangular form of $\ell^\pm$, so that
$\cB(0)=\ell^{+}\,(\ell^-)^{-1}$ is just a (invertible) triangular
decomposition\footnote{However, one has to allow square roots and
inverses for some generators.}. Thus, we conclude
that $\cB(0)$ generates the finite dimensional $\cU_{q}(gl(\enne-\emme))$
algebra.

One can construct from $\cB(z)$ another subalgebra in the following way.
One introduces
\begin{equation}
    \wh{\cB}(z)=\frac{1}{z^2}\cB^{-1}(z^{-1})\,,
\end{equation}
which satisfies:
\begin{eqnarray}
R_{12}(\frac{z}{w})\,\cB_{1}(z)\,R_{21}(zw)\,\wh{\cB}_{2}(w) =
\wh{\cB}_{2}(w)\,R_{12}(zw)\,\cB_{1}(z)\,R_{21}(\frac{z}{w})
\label{tra}\\
R_{12}(\frac{z}{w})\,\wh{\cB}_{1}(z)\,R_{21}(zw)\,\wh{\cB}_{2}(w) =
\wh{\cB}_{2}(w)\,R_{12}(zw)\,\wh{\cB}_{1}(z)\,R_{21}(\frac{z}{w})
\,.
\label{lala}
\end{eqnarray}
$\wh{\cB}(z)$ admits a $z^{-2}$ series expansion, so that we
get from (\ref{lala}) when $w\to\infty$:
\begin{eqnarray}
R_{21}^{-1}\,\wh{\cB}_{1}(z)\,R_{21}\,\wh{\cB}_{2}(\infty) =
\wh{\cB}_{2}(\infty)\,R_{12}\,\wh{\cB}_{1}(z)\,R_{12}^{-1}
\,.
\label{CR:Bt-infini}
\end{eqnarray}
Once again, it is the form of
$\widetilde{K}(z)=\frac{1}{z^2}\,K(z^{-1})^{-1}$ which determines
the subalgebra generated by $\wh{\cB}(\infty)$:
\begin{equation}
\widetilde{K}(\infty)
=diag(\underbrace{1,\ldots,1}_{\emme},
\underbrace{0,\ldots,0}_{\enne-\emme})\,.
\end{equation}
This proves that $\wh{\cB}(\infty)$
generates a $\cU_{q}(gl(\emme))$ algebra based on the generators
$\ell^{+}_{ij}$ and $\ell^{-}_{ij}$
with $i,j\leq\emme$.

Finally, since $\cB(0)$ (resp. $\wh{\cB}(\infty)$) depends on
$\ell^{+}_{ij}$ and $\ell^{-}_{ij}$ with $i,j>\emme$ (resp.
with $i,j\leq\emme$) only, we deduce:
\begin{lemma}\label{lem:Bo+Binf}
When $K(z)$ is diagonal, $\cB(0)$ (resp. $\wh{\cB}(\infty)$)
 generate a finite dimensional $\cU_{q}(gl(\enne-\emme))$ (resp.
$\cU_{q}(gl(\emme))$)
subalgebra of the quantum reflection algebra.\\
These two subalgebras commute one with each other, and thus are in direct sum
in the quantum reflection algebra.
\end{lemma}

\subsubsection{Symmetry of the transfer matrix}

To identify the symmetry algebra, we need the following lemma:
\begin{lemma}\label{lem:tr1}
For any (operator valued) matrix $\cB(z)$, one has
\begin{equation}
tr_{1}\left(M_{1}\,R_{12}\,\cB_{1}(z)\,R_{12}^{-1}\right)
=tr\left(M\,\cB(z)\right)\,\II
=tr_{1}\left(M_{1}\,R_{21}^{-1}\,\cB_{1}(z)\,R_{21}\right)\,.
\end{equation}
\end{lemma}
\underline{Proof:} We first prove the first equality
\begin{eqnarray}
 tr_{1}\left(M_{1}\,R_{12}\,\cB_{1}(z)\,R_{12}^{-1}\right) &=&
 tr_{1}\left(\big(M_{1}\,R_{12}\,\cB_{1}(z)\big)^{t_{1}}\,
 \big(R_{12}^{-1}\big)^{t_{1}}\right)=
 tr_{1}\left(\cB^t_{1}(z)\,R^{t_{1}}_{12}\,M_{1}\,
 \big(R_{12}^{-1}\big)^{t_{1}}\right)\qquad\\
 &=& tr_{1}\left(\cB^t_{1}(z)\,M_{1}\right)
 =tr_{1}\left(M_{1}\,\cB_{1}(z)\right)
\end{eqnarray}
where we have used (\ref{Rinv-t-inv}).\\
For the second equality, one uses (\ref{truc}) to write
\begin{eqnarray}
tr_{1}\left(M_{1}\,R_{21}^{-1}\,\cB_{1}(z)\,R_{21}\right) &=&
tr_{1}\left(M_{1}\,V_{1}\,V_{2}\,R_{12}^{-1}\,V_{1}\,V_{2}\,\cB_{1}(z)
\,V_{1}\,V_{2}\,R_{12}\,V_{1}\,V_{2}\right) \\
&=&
tr_{1}\left(\big(V^t_{1}\,V_{2}\,R_{12}^{-1}\,V_{1}\,V_{2}
\,\cB_{1}(z)\big)^{t_{1}}
\,\big(V_{1}\,V_{2}\,R_{12}\,V_{1}\,V_{2}\big)^{t_{1}}\right) \\
&=&
tr_{1}\left(\cB^t_{1}(z)\,V^t_{1}\,V_{2}\,\big(R_{12}^{-1}\big)^{t_{1}}
\,M^{-1}_{1}\,R_{12}^{t_{1}}\,V^t_{1}\,V_{2}\right) \\
&=&
tr_{1}\left(\cB^t_{1}(z)\,V^t_{1}\,V_{2}
\,M^{-1}_{1}\,V^t_{1}\,V_{2}\right) =
tr_{1}\left(\cB^t_{1}(z)\,M_{1}\right)
 =tr_{1}\left(M_{1}\,\cB_{1}(z)\right)\qquad
\end{eqnarray}
We used  (\ref{Rinv-t-inv}) in the last steps.
\finproof

Now, {from} the relations (\ref{CR:B0}) and (\ref{tra}), one deduces
\begin{eqnarray}
tr_{1}\left(M_{1}R_{12}\,\cB_{1}(z)\,R_{12}^{-1}\right)\,{\cB}_{2}(0) =
{\cB}_{2}(0)\,tr_{1}\left(M_{1}R_{21}^{-1}\,\cB_{1}(z)\,R_{21}\right)
\\
tr_{1}\left(M_{1}R_{21}^{-1}\,{\cB}_{1}(z)\,R_{21}\right)
\,\wh{\cB}_{2}(\infty) =
\wh{\cB}_{2}(\infty)\,
tr_{1}\left(M_{1}R_{12}\,{\cB}_{1}(z)\,R_{12}^{-1}\right)\,.
\end{eqnarray}
Thus, one gets the following property:
\begin{property}
When $K^{+}(z)=M$ and $K(z)$ is a diagonal matrix (\ref{eq:Kdiag}),
the transfer matrix $b(z)$ defined in (\ref{transferSP}) obeys
\begin{equation}
{[\cB(0)\,,\,b(z)]}={[\wh \cB(\infty)\,,\,b(z)]}=0\,.
\end{equation}
Hence, the open spin chain admits a finite dimensional $\cU_{q}(gl(\enne-\emme))\oplus
\cU_{q}(gl(\emme))$ symmetry algebra.
\end{property}

This property is valid whatever the quantum spaces are, and generalizes
the results obtained in \cite{mnsymm,done2} in the case of fundamental
representations.

 When $K^{+}(z)$ is a more general diagonal matrix, the symmetry is
reduced. To study this symmetry, we need the following lemma
\begin{lemma}
For a general diagonal matrix
$K^{+}(z)=diag(k_{1}(z),\ldots,k_{\enne}(z))$,
and for any (operator valued) matrix 
$\cB(z)=\sum_{a,b=1}^\enne B_{ab}(w)\,E_{ab}$,
one has:
\begin{eqnarray}
tr_{+}(w) &\equiv& tr_{2}\big(K^{+}_{2}(w)\,R_{12}^{-1}\,\cB_{2}(w)\,R_{12}\big) 
=b(w)\,\II + \sum_{1\leq a<b\leq\cN} x_{ab}(w)\,B_{ab}(w)\,E_{ab}
\qquad\label{def:t+}\\
tr_{-}(w) &\equiv& tr_{2}\big(K^{+}_{2}(w)\,R_{21}^{-1}\,\cB_{2}(w)\,R_{21}\big)
= b(w)\,\II + \sum_{1\leq a<b\leq\cN} x_{ab}(w)\,B_{ba}(w)\,E_{ba}
\qquad\label{def:t-}\\
x_{ab}(w) &=& (q-q^{-1})\Big(
q^{-1}\,k_{a}-q\,k_{b}-(q-q^{-1})\sum_{a<c<b}k_{c} \Big)
\label{def:xab}
\end{eqnarray}
where $b(z)=tr\big(K^{+}(z)\,\cB(z)\big)$.
\end{lemma}
\underline{Proof:} Direct calculation.\finproof

Then, one has:
\begin{property}
Let $K(z)$ ($K_{+}(z)$ resp.) be a diagonal solution of (\ref{re})  
(of (\ref{red}) resp.) with parameters $(\xi,\cM)$
(parameters $(\xi_{+},\cM_{+})$ resp.). One has the following
commutation relations:
\begin{eqnarray}
{[B_{ij}(0)\,,\,b(w)]} &=&0\ ,\qquad i,j\leq\cM_{+} \label{comBij0}\\
{[\wh B_{ij}(\infty)\,,\,b(w)]} &=&0\ ,\qquad i,j>\cM_{+}
\label{comBij-infty}
\end{eqnarray}
Hence, for
$\cM_{ax}=max(\cM,\cM_{+})$ and $\cM_{in}=min(\cM,\cM_{+})$, the open 
spin chain transfer matrix admits a
$\cU_{q}(gl(\cM_{in}))\oplus\cU_{q}(gl(\cM_{ax}-\cM_{in}))\oplus
\cU_{q}(gl(\cN-\cM_{ax}))$ symmetry algebra.
\end{property}
\underline{Proof:}
Starting with the reflection equation (\ref{re}), taking the limit
$z\to0$, then multiplying by $K^{+}_{2}(w)$ and taking the trace in
the space 2, one gets
$$\cB(0)\,tr_{+}(w) = tr_{-}(w)\,\cB(0)\,, $$
where we used the
notation (\ref{def:t+}) and (\ref{def:t-}).
The projection of this equation on $E_{ij}$ gives 
\begin{eqnarray}
{[B_{ij}(0)\,,\,b(w)]} &=&  \sum_{1\leq a<j}
x_{aj}(w)\, B_{ia}(0)\,B_{aj}(w) 
-\sum_{1\leq a<i} x_{ai}(w)\, B_{ia}(w)\,B_{aj}(0) \qquad
\label{eq:aux0}
\end{eqnarray}
where $x_{ab}(w)$ is defined in (\ref{def:xab}). 

Now, since $K^{+}(z)$ is a diagonal solution (\ref{diagKp}), its diagonal
terms obey 
\begin{equation}
k_{a}(z)=q^{2(b-a)}k_{b}(z) \quad\mbox{when}\quad a,b\leq\cM_{+}
\ \mbox{ or }\ a,b>\cM_{+}. 
\end{equation}
Using this property, one is led to $x_{ab}(z)=0$ 
when $a,b\leq\cM_{+}$ or when
$a,b>\cM_{+}$. Plugging this result in eq. (\ref{eq:aux0}) for
$i,j\leq\cM_{+}$, one gets
(\ref{comBij0}).

In the same way, starting with (\ref{tra}), taking the limit
$w\to\infty$, then exchanging the spaces 1 and 2, one obtains
$$\wh\cB(\infty)\,tr_{-}(w) = tr_{+}(w)\,\wh\cB(\infty)\,, $$
which leads to
\begin{eqnarray}
{[\wh B_{ij}(\infty)\,,\,b(w)]} &=&  \sum_{i<a\leq \cN}
x_{ia}(w)\,B_{ia}(w) \, \wh B_{aj}(\infty)
-\sum_{j<a\leq \cN} x_{ja}(w)\,\wh B_{ia}(\infty)\, B_{aj}(w) \qquad.
\label{eq:aux-infty}\nonumber
\end{eqnarray}
Hence, for $i,j>\cM_{+}$, one gets (\ref{comBij-infty}).

Finally, one concludes using the results of section \ref{sect:finiteSalg}
and considering the cases $i,j\leq\cM_{in}$,
$\cM_{in}<i,j\leq\cM_{ax}$ or $\cM_{ax}<i,j$.
\finproof

The implementation of more general non-diagonal boundaries reduces even
more the symmetry leading to a finite set of conserved
quantities that commute with the open transfer matrix, that is the boundary
quantum algebra \cite{doikoub, doikouh}.

\subsubsection{Finite dimensional representations \label{sect:reflRep}}

For the study of the representations of the reflection algebra, we
follow essentially the lines given in \cite{momo} for the
reflection algebra based on the Yangian of $gl(\enne)$, and in
\cite{MNO,molevbook,twmolev} for the twisted
Yangian.

We introduce the quantum comatrix, defined by
\begin{equation}
\wt \cT(z)\,\cT(zq^{\enne-1})=qdet \cT(z)\,,
\end{equation}
where the quantum determinant $qdet\cT(z)$, which generates the center
of $\cU_{q}(\wh{gl}(\enne))$, is defined in equation (\ref{def:qdet}).
The quantum comatrix is essentially the inverse of $\cT(z)$, which
motivates its use in the study of quantum reflection algebras.
It can be shown that it takes the form
\begin{eqnarray*}
&&\wt T_{ij}(z)= \\
&&\hspace{-1.2em}
\sum_{\sigma\in S_{\enne -1}} (-q)^{-\ell(\sigma)+j-i}\,
T_{a_{\sigma(1)},1}(z) T_{a_{\sigma(2)},2}(zq)\cdots
T_{a_{\sigma(i-1)},i-1}(zq^{i-2})T_{a_{\sigma(i)},i+1}(zq^{i-1})
\cdots T_{a_{\sigma(\enne -1)},\enne }(zq^{\enne -2})
\end{eqnarray*}
where $(a_{1},a_{2},\ldots,a_{\enne -1})=(1,\ldots,j-1,j+1,\ldots,\enne )$.
Moreover, it obeys the following lemma, proved using the steps
given in \cite{momo} for the reflection algebra (based on the Yangian of
$gl(\enne )$) and in \cite{twmolev,molevbook} for the twisted Yangians:
\begin{lemma}\label{lem:vacMono}
If $\omega$ is a highest weight vector of $\cT(z)$, with
eigenvalue $(P_{1}(z),\ldots,P_{\enne}(z))$, then it is also
a highest weight vector for the comatrix, with
\begin{equation}
\wt T_{ii}\,\omega=
P_{1}(zq^{\enne-2})\cdots P_{i-1}(zq^{\enne-i})
P_{i+1}(zq^{\enne-i-1})\cdots P_{\enne}(z)\,\omega
\equiv \wt P_{i}(z)\,\omega
\end{equation}
\end{lemma}
Since the quantum comatrix $\wt T(z)$ and the inverse matrix
$T(z)^{-1}$ are related through the quantum determinant, this lemma
also proves that $\omega$ is a highest weight vector for $T(z)^{-1}$,
with eigenvalue
\begin{equation}
T^{-1}(z)_{ii}\,\omega=\Big(qdet
T(zq^{-\enne+1})\Big)^{-1}\wt P_{i}(zq^{-\enne+1})\,\omega=
\left(\prod_{a=1}^{i-1}\frac{P_{a}(zq^{-a})}{P_{a}(zq^{-a+1})}\right)\,
\frac{1}{P_{i}(zq^{-i+1})}\,\omega
\equiv\wh P_{i}(z)\,\omega\,.
\label{val:Phat}
\end{equation}
This property is essential for the following theorem:
\begin{theorem}\label{theo:valB}
If $\omega$ is a highest weight vector of $\cT(z)$, with eigenvalue
$( P_{1}(z),\ldots, P_{\enne}(z))$, then, when $K(z)$ is a diagonal
matrix (\ref{eq:Kdiag}), it is also a highest weight vector for
$\cB(z)$, with eigenvalues
\begin{eqnarray}
B_{ii}(z)\,\omega &=&
\left(\Gamma_{i}(z)\,P_{i}(z)\,\wh P_{i}(z^{-1})
+ \,\sum_{k=1}^{i-1}\frac{q^{2}-1}{q^{2k}z^{4}-1}\,\Gamma_{k}(z)\,P_{k}(z)\,\wh P_{k}(z^{-1})
\right)\,\omega
\label{valBii}\\[1.2ex]
\Gamma_{k}(z)&=&
\frac{q^{k-1}\,\fb(z^2)}{\fb(q^{k-1}\,z^2)}\ \times
\left\{\begin{array}{lr} \alpha(z;\xi) &\quad;\quad k\leq\cM\\[1.2ex]
q^{-2\emme}\,\beta(zq^\emme;\xi) & \quad;\quad k>\cM
\end{array}\right.
\label{valGammaj}
\end{eqnarray}
where $\fb(z)=z-z^{-1}$, and
the expression of $\wh P_{j}(z)$ in terms of $P_{k}(z)$ is given
in (\ref{val:Phat}).
\end{theorem}
\underline{Proof:} One needs to compute $B_{ij}(z)\,\omega$ for
$j\leq i$. Denoting by $D_{k}(z)$ the diagonal elements of $K(z)$,
it takes the form:
\begin{eqnarray}
B_{ij}(z)\,\omega &=&
\sum_{1\leq k\leq j}T_{ik}(z)D_{k}(z)T^{-1}_{kj}(z^{-1})\,\omega
\label{Bom}\\
&=& \delta_{ij}\,D_{i}(z)T_{ii}(z)T^{-1}_{ii}(z^{-1})\,\omega
+\sum_{1\leq k\leq j} D_{k}(z)\,\Big[T_{ik}(z)\,,\,T^{-1}_{kj}(z^{-1})\Big]
\,\omega \nonumber
\end{eqnarray}
where $T^{-1}_{kj}(w)$ stands for $\big(\cT^{-1}(w)\big)_{kj}$.
{From} the relations (\ref{rtt}) applied to $\cT(z)$, one deduces (for
$k\leq j\leq i$):
\begin{eqnarray}
{\mathfrak b}(\frac{z}{w})\,\Big[T^{-1}_{kj}(w)\,,\,T_{ik}(z)\Big] &=&
\Big({\mathfrak a}(\frac{z}{w})-{\mathfrak b}(\frac{z}{w})\Big)\,\Big\{T_{ik}(z)T^{-1}_{kj}(w)-
\delta_{ij}T^{-1}_{kj}(w)T_{ik}(z)\Big\}
\nonumber\\
&&+(q-q^{-1})\,\left\{\frac{z}{w}\Big(\sum_{1\leq a<k}T_{ia}(z)T^{-1}_{aj}(w)-
\delta_{ij}\sum_{1\leq
a<j}T^{-1}_{ka}(w)T_{ak}(z)\Big)\right.
\qquad\nonumber\\
&&\qquad\quad
+\left.\frac{w}{z}\Big(\sum_{k<a\leq \enne}T_{ia}(z)T^{-1}_{aj}(w)-
\delta_{ij}\sum_{j<a\leq\enne}T^{-1}_{ka}(w)T_{ak}(z)\Big)\right\}
\label{comTTinv}
\end{eqnarray}
with ${\mathfrak a}(x)=qx-(qx)^{-1}$ and ${\mathfrak b}(x)=x-x^{-1}$,
see appendix \ref{sect:Rmatrix}.\\
\textit{We first consider the case $j<i$.}\\
Applying the above relation to $\omega$, one gets
\begin{equation}
\frac{-{\mathfrak a}({z}/{w})}{q-q^{-1}}\
T_{ik}(z)T^{-1}_{kj}(w)\,\omega =
\frac{z}{w}\sum_{1\leq a<k}T_{ia}(z)T^{-1}_{aj}(w)\,\omega+
\frac{w}{z}\sum_{k<a\leq \enne}T_{ia}(z)T^{-1}_{aj}(w)\,\omega
\end{equation}
Considering the case $k=j$, one obtains
\begin{equation}\sum_{1\leq a<j}T_{ia}(z)T^{-1}_{aj}(w)\,\omega=0\,.\end{equation}
Plugging this result in the former equation, we get, through iteration
\begin{equation}T_{ik}(z)T^{-1}_{kj}(w)\,\omega=0\,,\ k\leq j<i\,,\end{equation}
which proves that
\begin{equation}
B_{ij}(z)\,\omega=0\,,\ j<i\,.
\end{equation}
\textit{We now turn to the case $i=j$.}\\
Applying $\omega$ to the commutator (\ref{comTTinv}), one is led to
the following equations (for $k<i$):
\begin{eqnarray}
&&\frac{{\mathfrak a}(z/w)}{q-1/q} F_{ik}(z,w) +
\frac{z}{w}\sum_{a<k}F_{ia}(z,w) +\frac{w}{z}\left\{\sum_{k<a< i}
F_{ia}(z,w)-\sum_{a<k}G_{ka}(z,w)\right\} \\
&&\qquad=-
\frac{w}{z}\Big(\Psi_{i}(z,w)-\Psi_{k}(z,w)\Big)\\
&&\frac{{\mathfrak a}(z/w)}{q-1/q} G_{ik}(z,w) +
\frac{w}{z}\sum_{a<k}G_{ia}(z,w) +\frac{z}{w}\left\{\sum_{k<a< i}
G_{ia}(z,w)-\sum_{a<k}F_{ka}(z,w)\right\} \\
&&\qquad=-
\frac{z}{w}\Big(\Psi_{i}(z,w)-\Psi_{k}(z,w)\Big)
\end{eqnarray}
with
\begin{eqnarray}
F_{ik}(z,w)=T_{ik}(z)\,T^{-1}_{ki}(w)\omega\quad;\quad
G_{ik}(z,w)=T^{-1}_{ik}(w)\,T_{ki}(z)\omega\\
\Psi_{i}(z,w)=F_{ii}(z,w)=G_{ii}(z,w)=T_{ii}(z)T^{-1}_{ii}(w)\omega
= P_{i}(z)\wh P_{i}(w)\,\omega
\end{eqnarray}
This system in $F_{ab}(z,w)$ and $G_{ab}(z,w)$, $1\leq b<a\leq
\enne$, is triangular in the first index of $F$ and $G$, and
invertible (modulo lower indices) in the highest first index, so
that it admits a unique solution. It is then easy (but lengthy) to
show that the solution is
\begin{eqnarray}
    F_{ik}(z,w) &=& (q^2-1)\,
    F_{ik}^{red}(z,w)\,\\
   G_{ik}(z,w) &=& (q^2-1)\,\left(\frac{z}{w}\right)^2\,q^{2k-2}\,
    G_{ik}^{red}(z,w)
\end{eqnarray}
where $F^{red}$ et $G^{red}$ are linear combinations of the  $\Psi$'s:
\begin{eqnarray*}
 F^{red}_{ik}(z,w) &=& \frac{\Psi_{k}(z,w)}{q^{2k}x^2-1}
 -\frac{q^{2i-2k-2}\,\Psi_{i}(z,w)}{q^{2i-2}x^2-1}
 +(1-q^2)\sum_{a=k+1}^{i-1}\frac{q^{2(a-k-1)}\,
 \Psi_{a}(z,w)}{(q^{2a}x^2-1)(q^{2a-2}x^2-1)}\quad\\
 G^{red}_{ik}(z,w) &=& \frac{\Psi_{k}(z,w)}{q^{2k}x^2-1}
 -\frac{\Psi_{i}(z,w)}{q^{2i-2}x^2-1}+
 (1-q^2)\sum_{a=k+1}^{i-1}\frac{q^{2a-2}\,x^2\,
 \Psi_{a}(z,w)}{(q^{2a}x^2-1)(q^{2a-2}x^2-1)}
 \end{eqnarray*}
where $x=\frac{z}{w}$. Since $\Psi_{k}(z,w)$ is proportional to
$\omega$, this proves that $\omega$ is a highest weight for $\cB(z)$.
Plugging the value of $F_{ik}(z,z^{-1})$ into the equation (\ref{Bom}),
one gets the eigenvalues
\begin{eqnarray}
B_{ii}(z)\,\omega &=&
\left\{D_{i}(z)\,P_{i}(z)\,\wh P_{i}(z^{-1})
+ (q^{2}-1)\,\sum_{1\leq
k<i}D_{k}(z)\left(\frac{P_{k}(z)\,\wh P_{k}(z^{-1})}{q^{2k}z^{4}-1}
-q^{2(i-k-1)}\,\frac{P_{i}(z)\,\wh P_{i}(z^{-1})}{q^{2i-2}z^{4}-1}\right.\right.
\nonu
&& \qquad\qquad
\left.\left. +(1-q^2)\sum_{a=k+1}^{i-1} q^{2(a-k-1)}\,
\frac{P_{a}(z)\,\wh P_{a}(z^{-1})}{(q^{2a}z^{4}-1)(q^{2a-2}z^{4}-1)}
\right)\right\}\,\omega\,,
\end{eqnarray}
which can be rewritten in the form
\begin{eqnarray}
B_{ii}(z)\,\omega &=&
\left\{\Gamma_{i}(z)\,P_{i}(z)\,\wh P_{i}(z^{-1})
+ \,\sum_{k=1}^{i-1}\frac{q^{2}-1}{q^{2k}z^{4}-1}\,\Gamma_{k}(z)\,P_{k}(z)\,\wh P_{k}(z^{-1})
\right\}\,\omega
\nonu
\Gamma_{k}(z)&=& D_k(z)
-\frac{q^2-1}{q^{2k-2}z^{4}-1}\,\sum_{a=1}^{k-1} q^{2k-2a-2}\,
D_{a}(z)
\end{eqnarray}
Implementing the diagonal form (\ref{eq:Kdiag}), we get
(\ref{valBii})-(\ref{valGammaj}).
\finproof
As a consequence of this theorem, we can compute the transfer matrix
eigenvalue of the pseudo-vacuum $\omega$:
\begin{corollary}\label{coro:valTransf}
When $K^{+}(z)$ is a diagonal matrix (\ref{diagKp}), the highest weight
vector $\omega$ is an eigenvector of the
transfer matrix $b(z)$ given in (\ref{transferSP}):
\begin{eqnarray}
b(z)\,\omega &=& \Lambda^{0}(z)\,\omega \qmbox{with}
\Lambda^{0}(z) =\sum_{j=1}^\enne g_{j}(z)\,P_{j}(z)\wh
P_{j}(z^{-1})
\label{lambda0-open}\\
g_{j}(z) &=&
\Gamma_{j}(z)\,q^{-j+1}\,
\frac{\fb(q^{\enne}\,z^2)}{\fb(q^{j}\,z^2)}\ \times
\left\{\begin{array}{lr}
q^{2\enne-2\emme_{+}}\,\wt\alpha(zq^{\emme_{+}-\enne}\,;\xi_{+})
&\quad;\quad j\leq\cM_{+}\\[1.2ex]
\wt\beta(z;\xi_{+}) & \quad;\quad j>\cM_{+}
\end{array}\right.
\label{eq:defg}
\end{eqnarray}
where $\Gamma_{j}(z)$ are given in (\ref{valGammaj}) and $\wh
P_{j}(z)$ in (\ref{val:Phat}).
\end{corollary}
\underline{Proof:} Direct calculation using the expression of $b(z)$, the
diagonal form (\ref{diagKp}) for $K^{+}(z)$ and the eigenvalues (\ref{valBii}).
\finproof
Although each function $g_j(z)$ possesses some poles at the points
$z^2=q^{-j}$, the residues of $\Lambda^{0}(z)$ vanish due to the
particular forms of $\wh P(z)$, $\alpha(z;\xi)$, $\beta(z;\xi)$,
$\wt \alpha(z;\xi_+)$ and $\wt \beta(z;\xi_+)$.

\subsubsection{Quantum contraction and Sklyanin determinant}

As in the case of the Yangian, we can construct series whose 
coefficients give central elements. In appendices \ref{sect:fQ}
and \ref{sect:gefu}, we recall the construction of two of them: the
quantum contraction of $\cB(z)$ and the Sklyanin determinant. In the
spin chain context, they allow one to obtain constraints on the
eigenvalues of the problem.\\
If $K(z)$ is given by relation
(\ref{eq:Kdiag}) and $K^+(z)$ by (\ref{diagKp}), then their quantum
contractions take the following simple forms
\begin{eqnarray}
\delta(K(z))=\theta_0q^{2{\cal M}-2 \cal N}\;
\alpha(z\rho;\xi)\;\beta(z\rho;\xi)\;\left(\frac{1}{z^2}-z^2\right)
\end{eqnarray}
and
\begin{eqnarray}
\delta(K^+(z))=\theta_0q^{2\cal M_+}
\wt{\alpha}\left(\frac{zq^{\emme_+}}{\rho
};\xi_+\right)\wt{\beta}\left(\frac{z\rho}{q^{\emme_+}};\xi_+\right)\;
\left(q^{2\cal N}z^2-\frac{1}{q^{2\cal N} z^2}\right)
\end{eqnarray}
where we recall that $\rho=q^{\enne/2}$. For a given representation
of the reflection algebra, the quantum contraction of the generators
is simply given by
$\delta(\cB(z))=\delta(K(z))$. \\
In the same way, the Sklyanin
determinant can be computed in a particular representation and
becomes
\begin{equation}
sdet(\cB(z))=sdet(K(z))\prod_{a=1}^\enne
\frac{P_a(zq^{\enne-a})}{P_a(z^{-1}q^{1-a})}\,. \label{eq:sdetB}
\end{equation}
The Sklyanin determinant of the diagonal solution
(\ref{eq:Kdiag}) takes the following value
\begin{equation}
sdet(K(z))=q^{2\emme(\emme-\enne)}\prod_{1\leq a \leq \emme}
\alpha(z_aq^{\enne-\emme}) \prod_{\emme+1\leq a \leq \enne}
\beta(z_a) \prod_{1\leq a< b \leq
\enne}\mathfrak{b}\left(\frac{q}{z_az_b}\right) \;,\label{eq:sdetKm}
\end{equation}
where we remind $\mathfrak{b}(z)=z-1/z$ and $z_a=zq^{a-1}$.
We can also compute the Sklyanin determinant of $K^+(z)$ when this
latter matrix is
given by relation (\ref{diagKp}):
\begin{equation}
sdet(K^+(z))=q^{2\emme_+(\enne-\emme_+)}\prod_{1\leq a \leq \emme_+}
\widetilde{\alpha}(z_a) \prod_{\emme_{+}+1\leq a \leq \enne}
\widetilde{\beta}(z_aq^{-\emme_+}) \prod_{1\leq a< b \leq
\enne}\mathfrak{b}\left(qz_az_b\rho^2 \right)\;. \label{eq:sdetKp}
\end{equation}
The computation of the previous explicit forms of the Sklyanin
determinant follows the lines of the proof made in the case of the
reflection algebra associated to the Yangian \cite{momo}.

\section{Spin chains\label{sect:spch}}

Having defined the underlying algebraic structures, we now turn to the
construction of spin chains. To each of the above algebras will be
associated a different boundary condition: periodic (closed spin
chain) or soliton preserving (SP open spin chain). In both cases, the monodromy
matrix will obey the defining relations of the corresponding algebra.

\subsection{The periodic spin chain\label{sect:closed}}

Let us first consider the algebra $\cU_{q}(\wh{gl}({\cal N}))$ as introduced in section
\ref{sect:Uq} and the corresponding
algebraic monodromy matrix  $\cT(z)=\cT^+(z)$, given in (\ref{mono}).

The transfer matrix of the system has been also defined (at the algebraic
level) in (\ref{transfer}), and the commutation relation (\ref{com})
 ensures the integrability of the model. We repeat that our description is
purely algebraic at this stage, i.e. the entailed results are independent
of the choice of representation, and thus are universal. Once we
assign particular representations on each of the copies of $\cU_{q}(\wh{gl}({\cal N}))$,
which then become the so-called `quantum spaces' of the spin chain, the algebraic
construction of the monodromy and transfer matrices acquires a physical meaning.
Then, one may diagonalize the transfer matrix
(\ref{transfer}), which is the quantity that encodes all the physical
information of the system, and derive the corresponding Bethe ansatz
equations \cite{faddeev}. This will be in fact our main objective in the
subsequent sections.

Let us also remark that the Borel subalgebra generated by $\cT^+(z)$
can be viewed as a deformation of $\cY_{\enne}$, the Yangian of
$gl(\enne)$, so that the present spin chains are `deformation' of
the spin chain models build on $\cY_{\enne}$. Hence the present
algebraic construction can be viewed as the `deformed' counterpart
of the one done for the Yangian in \cite{byebye}. The  study of spin
chains based on the full $\cU_{q}(\wh{gl}({\cal N}))$ algebra (or on
the quantum double of $\cY_{\enne}$) is still lacking (to our
knowledge). In the same way, we have assumed that the irreducible
representations are finite dimensional, to ensure the existence of
the Bethe ansatz reference state. However, one can also define spin
chains with infinite dimensional representations, although a Bethe
ansatz is lacking in this context. These spin chains are the ones
used in large $N$ QCD: it is clear that a  study of them would be of
interest. Note that in the case of infinite dimensional
representations, one can take a non-vanishing central charge $c$:
the different definitions for the monodromy matrix will then become
inequivalent.

\subsubsection{Spectrum of the periodic spin chain}

We shall now derive the spectrum of the periodic spin chain by implementing
the analytical Bethe ansatz formulation.
We denote by $T_{0}(z)$ the represented monodromy matrix.

The first step is to derive an appropriate reference state, that is
a state which is an eigenvector of the transfer matrix. This is
provided by the highest weight vector $\omega$ of the
$\cU_{q}(\wh{gl}({\cal N}))$ representations, as presented in section
\ref{Uqrep}. Its eigenvalue is related to the Drinfel'd
polynomials characterizing the
representation, see (\ref{lambda0}) and (\ref{polyDrin}).\\
Having determined the form of the pseudovacuum eigenvalue we may now assume
the following form for the general eigenvalue
\begin{equation}
  \Lambda(z) = \sum_{a=1}^{{\cal N}}\ P_{a}(z)\
  A_{a-1}(z)\,,\label{eq:Lambda-closed}
\end{equation}
where the dressing functions $A_{a}(z)$ may be derived by
implementing a number of constraints upon the spectrum (see e.g
\cite{resef, jim, mnanal, byebye} for a more detailed description). More
specifically, the constraints follow from:

\begin{enumerate}
  \item
  The fusion expressions (\ref{fusionQ}) produces
  constraints between the dressing functions of $t$ and those of $\wh t$
  (defined by relation (\ref{tr11})), while the
  generalized fusion (\ref{fusion2}) provides a relation among the 
  dressing functions of $t$.
  Explicitly, using relation (\ref{eq:qdetrep}), the constraint reads as
  \begin{equation}
  \prod_{a=1}^{\cal N} A_{a-1}(zq^{N-a})=1\;.
  \end{equation}
  \item
  Analyticity requirements, imposed on the spectrum, lead  any two
  successive (up to relabeling) dressing functions to have common poles.
  \item
  The fact that the $R$ matrix and monodromy matrix in the chosen
  representations are written in terms of
  rational functions (in $z$) leads to the assumption that $A_{a}$ should be
  given as products of rational functions.
  \item
  The asymptotic behaviour of the transfer matrix provides important
  information about the form of the aforementioned products (see
  section \ref{sec:cartclosed}).
  \item
  The parity of the transfer matrix\footnote{Using an additive spectral
  parameter  this parity is nothing but the periodicity
  $t(\lambda+\frac{i\pi}{\mu})=t(\lambda)$ where $\mu$ is related to
  the deformation parameter through $q=e^{i\mu}$ and the spectral
  parameters $z=e^{\mu\lambda}$.}
  \begin{equation}
    t(-z) = t(z),
  \end{equation}
  leads to the parity of the eigenvalues.
\end{enumerate}
The following set of dressing functions satisfy all the
aforementioned constraints, for $a=0,\ldots,\enne-1$,
\begin{eqnarray}
A_{a}(z) &=& \prod_{\ell=1}^{M^{(a)}} \frac{q^{a+2} \,{z}^2
-\big({z_{\ell}^{(a)}}\big)^{2} }
{q^{a+1}\,{z}^2-q\,\big({z_{\ell}^{(a)}}\big)^2 } \
\prod_{\ell=1}^{M^{(a+1)}} \frac{q^{a}\,{z}^2
-q\,\big({z_{\ell}^{(a+1)}}\big)^2 }
{q^{a+1}\,{z}^2-\big({z_{\ell}^{(a+1)}}\big)^2 }\;, \label{dress}
\end{eqnarray}
where, by convention, $M^{(0)}=0$ and $M^{(\enne)}=0$. We show in
the next section how the coefficients $M^{(a)}$ are related to the
eigenvalues of the Cartan's generators. Finally, requiring
analyticity of the spectrum, we obtain the following set of Bethe
ansatz equations, for $a=1,\ldots,\enne$ and $k=1,\ldots,M^{(a)}$:
\begin{eqnarray}
 \frac{P_{a}(q^{-\frac{a}{2}}\,z_{k}^{(a)}) }{
  P_{a+1}(q^{-\frac{a}{2}}\,z_{k}^{(a)})} = -\prod_{\ell=1}^{M^{(a-1)}}
  e_{-1}\left(\frac{z_{k}^{(a)} }{z_{\ell}^{(a-1)}}\right) \ \prod_{\ell=1}^{M^{(a)}}
  e_{2}\left(\frac{z_{k}^{(a)} }{z_{\ell}^{(a)}}\right)\ \prod_{\ell=1}^{M^{(a+1)}}
  e_{-1}\left(\frac{z_{k}^{(a)} }{z_{\ell}^{(a+1)}}\right)
\,,\
\label{BAE-closed}
\end{eqnarray}
where we defined
\begin{equation}
  e_{n}(z) = \frac{z^{2}\,q^n-1}{z^{2}-q^n}\;.
\end{equation}
Note that, due to the normalisations we have chosen, the transfer matrix is
analytical everywhere but zero. Hence, the above derivation is a
priori valid for $z_{k}^{(a)}\neq0$. However, since we are dealing
with finite dimensional representations, multiplying by an appropriate
power of $z$, one can always cure the
unphysical pole in zero.

Let us remark that the right-hand side of the BAE reflects the
Lie algebra dependence (through the
Cartan matrix of $gl(\cN)$), while their
left-hand side shows up a
representation dependence (it can be rewriten in terms of Drinfel'd
polynomials solely, using the relation (\ref{polyDrin})).
The choice of a closed spin chain model
is fixed by the choice of the quantum spaces, i.e. the choice
of the Drinfel'd polynomials which determine the values $P_{k}(\lambda)$. Once these polynomials
are given, the spectrum of the transfer matrix is fixed through the
resolution of the BAE.

\subsubsection{Dressing and Cartan generators \label{sec:cartclosed}}

We have seen that the Cartan generators of the finite dimensional
$\cU_{q}(gl(\enne))$ algebra commute with the transfer matrix. It is
thus natural to try to connect the dressing used for $\Lambda(z)$ to
the eigenvalues of the Cartan generators. It is done in the following
way.

One first take the the limit $z\to\infty$ to get
\begin{equation}
\Lambda(z) \
\raisebox{-1ex}{$\stackrel{\displaystyle
\sim}{\scriptscriptstyle{z\to\infty}}$}\  z^{N}\
\sum_{j=1}^\enne
q^{M^{(j-1)}-M^{(j)}} \prod_{n=1}^{N} q^{a_{n}+\varpi_{j}^{(n)}}
\end{equation}
where by convention $M^{(0)}=0$ and $M^{(\enne)}=0$. On the other
hand, since $\cL(z)\ \raisebox{-1ex}{$\stackrel{\displaystyle
\sim}{\scriptscriptstyle{z\to\infty}}$}\ z\,\cL^{+}$ and $\cL^{+}$
is triangular, one can also compute
\begin{equation}
t(z)\
\raisebox{-1ex}{$\stackrel{\displaystyle\sim}{{\scriptscriptstyle{z\to\infty}}}$}
\ z^{N}\ \sum_{j=1}^\enne
\underbrace{\ell_{jj} \otimes\ldots\otimes \ell_{jj}}_{N}
\end{equation}
Now, the Cartan generators $h_{j}^{(1\ldots N)}$, $j=1,\ldots,\enne$,
defined by
\begin{equation}
\underbrace{\ell_{jj} \otimes\ldots\otimes \ell_{jj}}_{N} =
q^{h^{(1)}_{jj} \oplus\ldots\oplus h^{(N)}_{jj}} = q^{h_{j}^{(1\ldots
N)}}\,,
\end{equation}
where the superscript indicates in which quantum space(s) acts the
operator, have been proved to commute with $t(z)$. Then, starting
from a transfer matrix eigenvector
\begin{equation}
t(z)\,v=\Lambda(z)\,v\ ,
\end{equation}
one can deduce its $h^{(1\ldots N)}_{j}$ eigenvalue:
\begin{equation}
h_{j}^{(1\ldots N)}\,v = \lambda_{j}\,v= \Big( M^{(j-1)}-M^{(j)}+\sum_{n=1}^{N}(
{a_{n}+\varpi_{j}^{(n)}})\Big)\ v\ ,\
\forall\,j=1,\ldots,\enne\quad\mbox{with } M^{(0)}=0\,.
\label{h-eigenvalue}
\end{equation}
This result generalizes the one obtained for the usual closed spin
chain (see e.g. \cite{done2,mnanal}). Indeed, in this latter case,
all the sites carry a fundamental representation, so that
$\varpi_{j}^{(n)}=\delta_{j,1}$, $\forall\ n=1,\ldots,N$. Then, for
the $sl(\enne)$ Cartan generators $s_{j}=h^{(1\ldots
N)}_{j+1}-h^{(1\ldots N)}_{j}$, one gets
\begin{equation}
s_{j}\,v = \Big( 2M^{(j)}-M^{(j-1)}-M^{(j+1)}-N\,\delta_{j,1}\Big)\ v\ ,
\ \forall\,j=1,\ldots,\enne-1\quad\mbox{with } M^{(0)}=0\,.
\end{equation}
It explains the usual convention $M^{(0)}=N$ used generally for this
spin chain. However, from the general result (\ref{h-eigenvalue}), we
rather
use the convention $M^{(0)}=0$, which is more natural and allows a
more condensed writing of the BAE.

\subsubsection{Upper bounds on $M^{(j)}$\label{M-borne}}
We consider here the case where $q$ is not root of unity, where we
have the following theorem:
\begin{theorem}\cite{rosso}\\
\label{theo:repUq}
The one-dimensional irreducible representations of $\cU_{q}(gl(\enne))$
have highest weight $\bs{\eta}=(\eta_{1},\ldots,\eta_{\enne})$, 
with $\eta_{j}\,,j=1,\ldots,\enne$ taking values in $\{1,i,-i,-1\}$.\\
Any finite dimensional irreducible representation of
$\cU_{q}(gl(\enne))$ is described by an highest weight
$\bs{\wt\varpi}=\bs{\eta}\cdot q^{\bs \varpi}$, where $\bs{\varpi}$ is a
dominant weight of $gl(\enne)$ and $\bs{\eta}$ describes a
one-dimensional representation.
\end{theorem}
Then, up to the one-dimensional representations, the study of (finite 
dimensional) representations of $\cU_{q}(gl(\enne))$ is 
equivalent to the case of classical Lie algebra. For simplicity, we will assume in
this section that the representations in each sites have
$\eta_{j}=1$, $\forall j$. Then the tensor product of such
representations closes on representations of same type.

The values (\ref{h-eigenvalue}) allow us to recover the upper bounds
for the parameters $M^{(j)}$. Indeed, the Bethe ansatz hypothesis
states that the highest weight vectors of the
diagonal $gl(\enne)$ algebra, whose Cartan basis is spanned by the
$h_{j}^{(1\ldots N)}$ generators, are transfer matrix  eigenvectors
$v$, and  moreover that the eigenvalues of these eigenvectors span the 
whole set of transfer matrix eigenvalues. Then, the different dressings
are in one-to-one correspondance with the different $gl(\enne)$
representations entering the spin chain.
This also implies that
$\bs{\lambda}=(\lambda_{1},\ldots,\lambda_{\enne})$
given in (\ref{h-eigenvalue}) must be a dominant
weight for the $gl(\enne)$ algebra. 

The spin chain, as a representation of the
$gl(\enne)$ algebra, reads 
\begin{equation}
\bigotimes_{n=1}^{N}V(\bs{\varpi^{(n)}})
=\bigoplus_{\bs{\lambda}\leq\bs{\varpi}}
c_{\lambda}\,V(\bs{\lambda}) \label{eq:decomp-glN}
\end{equation}
{where} 
\begin{equation}
\bs{\varpi}=\sum_{n=1}^{N}\bs{\varpi^{(n)}}\,
\end{equation}
correspond to the pseudo-vacuum eigenvalue,
$c_{\lambda}$ are multiplicities and the sum is done on dominant
weights smaller than $\bs{\varpi}$. We remind that the partial order on
weights is defined as follow:  one has $\bs{\lambda}\leq\bs{\varpi}$ if and
only if $\bs{\varpi}-\bs{\lambda}$ is a positive root. Computing this
quantity for the eigenvalues (\ref{h-eigenvalue}), one gets
\begin{equation}
\bs{\varpi}-\bs{\lambda}=\sum_{k=1}^{\enne-1}
M^{(k)}\,(\eps_{k}-\eps_{k+1})\,,
\end{equation}
where $(\eps_{k}-\eps_{k+1})$, $k=1,\ldots,\enne-1$, are the $sl(\enne)$ simple roots.
Demanding
the weights obtained from (\ref{h-eigenvalue}) to be indeed smaller (in 
the above sense) than $\bs{\varpi}$, one recovers that 
the parameters $M^{(j)}$ are positive integers. 

{F}rom the decomposition (\ref{eq:decomp-glN}), demanding that the
weights $\bs\lambda$ are dominant, one can deduce upper
bounds on the integers $M^{(j)}$.
Since $\lambda_{j}$, $j=1,\ldots,\enne$
correspond to the decomposition of $\bs{\lambda}$ on the fundamental
weights $\eps_{j}$, one must have $\lambda_{j}-\lambda_{j+1}\in\ZZ_{+}$, $\forall j$.
Computing this quantity, one gets
\begin{equation}
2M^{(j)}- M^{(j-1)} - M^{(j+1)}\leq n_{j}
\ \mbox{ with } n_{j}=\varpi_{j}-\varpi_{j+1} \ \mbox{ and }
\varpi_{j}=\sum_{n=1}^{N}\varpi_{j}^{(n)}\,,\ 
j=1,\ldots,\enne. \label{eq:Mj-brute}
\end{equation}
Note that $n_{j}\in\ZZ_{+}$ because we supposed as a starting point
that $\bs\varpi$ defines an irreducible finite dimensional
representation of $gl(\enne)$.

The condition (\ref{eq:Mj-brute}) rewrites as 
\begin{equation}
(A\,\vec M)_{j}\leq n_{j} \ \mbox{ with } A \mbox{ the $sl(\enne)$-Cartan matrix and
} \vec M=\left(\begin{array}{c} M^{(1)} \\ \vdots \\ M^{(\enne-1)}
\end{array}\right)\ .
\end{equation}
Using the fact that $A^{-1}$ has only positive entries, one deduces
\begin{equation}
M_{j}\leq (A^{-1}\,\vec n)_{j} \qmbox{with} 
\vec n=\left(\begin{array}{c} n_{1}\\ \vdots \\ n_{\enne-1}
\end{array}\right)\ .
\end{equation}
{F}rom the explicit form $\enne\,(A^{-1})_{ij}=min(i,j)\,\big(\enne-max(i,j)\big)$,
one gets 
\begin{equation}
\enne\, M_{j}\leq \sum_{k=1}^j k(\enne-j)\,n_{k} +\sum_{k=j+1}^{\enne-1}
j(\enne-k)\,n_{k}\ ,\ \forall j=1,\ldots,\enne-1\ .
\end{equation}
In the particular case of fundamental representations
($\varpi_{j}^{(n)}=\delta_{j,1}$), one has $\varpi_{k}=N\delta_{k,1}$, 
so that one recovers the
condition
\begin{equation}
M^{(j)} \leq \frac{\enne-j}{\enne}\, N\ .
\end{equation}

\subsubsection{Example: closed spin chain with one defect
\label{example-closed}}
As an example, we study the case of an arbitrary defect in a fundamental
spin chain. In term of representations, it means that all the
representations (all the sites of the spin chain) but one are taken to be the
fundamental representation of $\cU_{q}(gl(\enne))$. 
We suppose this particular site to be the site
$p$, with $1<p<N$. 

The $\cU_{q}(gl(\enne))$ highest weights on each site then read
$\varpi^{(n)}_{a}=\delta_{a,1}$ for $n\neq p$, so that the Cartan eigenvalues
simplify to
\begin{equation}
 P_{a}(z) =\left\{\begin{array}{ll}
 (zq-z^{-1}q^{-1})^{N-1}\, (zq^{\varpi_{1}+\theta}-z^{-1}q^{-\varpi_{1}-\theta})
 &\qmbox{if} a=1 \\[1.2ex]
  (z-z^{-1})^{N-1}\, (zq^{\varpi_{a}+\theta}-z^{-1}q^{-\varpi_{a}-\theta})
 &\qmbox{if} a\neq1
 \end{array}\right.
 \end{equation}
where, for simplicity, we have dropped the superscript $p$ on the
highest weight ${\bs\varpi}^{(p)}$ and set all the inhomogeneity
parameters to 0, except the one of site $p$, that we noted $\theta$.

The monodromy matrix reads
\begin{equation}
T_{0}(z) = R_{01}(z)\cdots R_{0,p-1}(z)\,L_{0p}(zq^\theta)\,R_{0,p+1}(z)
\cdots R_{0N}(z)
\end{equation}
so that one can define a local Hamiltonian (prime denotes derivative
w.r.t. $z$):
\begin{eqnarray}
 \cH &=& t(1)^{-1}\,t'(1)=\cH_{p-1,p,p+1} +\sum_{\atopn{j=1}{j\neq
 p,p-1}}^{N} \cH_{j,j+1} \label{eq:Hclosed1}
 \\
 \cH_{j,j+1}&=&\frac{1}{q-q^{-1}}\,\cP_{j,j+1}\,R'_{j+1,j}\quad j\neq
 p\,,\,p-1
 \\[1.2ex]
 \cH_{p-1,p,p+1}&=&L^{-1}_{p+1,p}(q^\theta)\,L'_{p+1,p}(q^\theta)+
 L^{-1}_{p+1,p}(q^\theta)\,\cP_{p-1,p+1}\,R'_{p-1,p+1}\,L_{p+1,p}(q^\theta)
 \qquad\label{eq:Hclosed3}\\[1.2ex]
  R'_{ab}&=&2\big(\II-P_{ab}^q\big)+(q+q^{-1})\,\cP_{ab}\,,\quad
  \forall\ a\neq b.
\end{eqnarray}
 This formula is a direct calculation from $t(z)=tr_{0} T_{0}(z)$, using the
 properties of ($q$-)permutations and the value of
 $R(1)=(q-q^{-1})\cP$ and $ R'(1)$.
 As usual in periodic spin
 chains, we have identified the site $N+1$ with the site $1$.\\
 The energies are of the form $E=\Lambda'(1)/\Lambda(1)$, where
 $\Lambda(z)$ is given in (\ref{eq:Lambda-closed}), with dressings
 (\ref{dress}), BAEs (\ref{BAE-closed}) and $P_{a}(z)$ as
 above. A straightforward calculation leads to
\begin{eqnarray}
    E &=& E_{0}+ 2\,(q^2-1)\,\sum_{\ell=1}^{M^{(1)}}
\frac{\big(z_{\ell}^{(1)}\big)^2}{ \left(1-q\,\big(z_{\ell}^{(1)}\big)^2\right)
\left(q-\big(z_{\ell}^{(1)}\big)^2\right)}\\[1.2ex]
E_{0} &=& (N-1)\,\frac{q+q^{-1}}{q-q^{-1}}+
\frac{q^{\varpi_{1}+\theta}+q^{-\varpi_{1}-\theta}}
{q^{\varpi_{1}+\theta}-q^{-\varpi_{1}-\theta}}
\label{E-pseudo}
\end{eqnarray}
Here, $E_{0}$ is the energy of the pseudo-vacuum, with 
normalisations as given in (\ref{eq:Hclosed1})-(\ref{eq:Hclosed3}).\\
Of course, when the `defect' representation is also the fundamental
representation, one recovers the usual XXZ model.

\subsection{Soliton preserving open spin chain\label{sect:open}}

This section is devoted to the derivation of the spectrum and Bethe ansatz
equations for the integrable open spin chain with soliton preserving
diagonal boundaries. These boundary conditions 
physically describe the reflection of a soliton to
itself, i.e. no multiplet change occurs under reflection. The
algebraic structure associated to this kind of open spin chain is the
quantum reflection algebra described in section \ref{sect:qrefl}.

Our main aim consists in building the corresponding quantum system,
i.e.  the open
quantum spin chain. The open spin chain may be constructed following the
generalized QISM, introduced by Sklyanin \cite{sklyanin}. It relies on
 tensor product realizations of the general solution
(\ref{gensol}), on the open spin chain
monodromy matrix
\begin{equation}
  {\cal B}_{0}(z) = \cT_{0}(z)\ K_{0}^{-}(z)\ \cT^{-1}_{0}(z^{-1}),
  \label{transfer0}
\end{equation}
and on the transfer matrix
$b(z)=tr_{0}\left(K^{+}_{0}(z)\,\cB_{0}(z)\right)$, as they were
introduced in section \ref{sect:defOpen}.

\subsubsection{Spectrum of the open spin chain}

As in the closed case, our ultimate goal is to derive the spectrum and the
corresponding Bethe ansatz equations. To achieve that we shall need an
appropriate reference state. Fortunately, if we restrict our attention 
to
the case where both left and right boundaries are diagonal,  there
exists an obvious reference state: the  highest weight
vector of the $\cU_{q}(\wh{gl}({\cal N}))$ representation (see section
\ref{sect:reflRep}).

Having specified the form of the pseudo-vacuum
eigenvalue (\ref{lambda0-open}),
we can make the following assumption for the general eigenvalue
of the open transfer matrix:
\begin{equation}
  \Lambda(z) =\sum_{a=1}^{{\cal N}}  g_{a}(z)\
  P_{a}(z)\ \wh P_{a}(z^{-1})\ \wt A_{a-1}(z)\,.
  \label{eq:Lambda-open}
\end{equation}
The dressing functions $\wt A(z)$ may be derived by
implementing a number of constraints upon the spectrum (see e.g
\cite{doikou2, selene, byebye} for a more detailed description). In
particular, similarly to the periodic case the constraints follow
from:
\begin{enumerate}
\item The fusion relation (\ref{fusion3}). Using the explicit form
of the Sklyanin determinant (\ref{eq:sdetB})-(\ref{eq:sdetKp}) and
the value (\ref{eq:defg}) of the function $g_a(z)$, we find the
following constraint
\begin{equation}
  \prod_{a=1}^{\cal N} \wt A_{a-1}(zq^{N-a})=1\;.
\end{equation}
Note that this constraint is similar to the one found in the
periodic case and does not depend on the chosen representations and
on the boundaries.
\item Parity requirements which means that the dressing functions
are only function of $z^2$.
\item The fact that the $R$ matrix and monodromy matrix (in the chosen
representations) are written in terms of rational functions which
implies that $\wt A(z)$ are rational functions.
\item The remark that every two successive $g_{a}$ have common poles, which
must vanish to ensure analyticity of the eigenvalues. Namely, the
vanishing of the residues at $z^2=q^{-a}$ ($1\leq a \leq \enne-1$)
implies the following constraints
\begin{equation}
\wt A_{a-1}(q^{-a/2})=\wt A_{a}(q^{-a/2})\;.
\end{equation}
\end{enumerate}
The asymptotic behaviour of the transfer matrix also provides
important information about the form of the aforementioned products
as explained in section \ref{sec:cartopen} (see also \cite{mnanal,
doikou2}). \\
The following set of dressing functions satisfy all
the aforementioned constraints:
\begin{eqnarray}
\wt A_{a}(z) &=& \prod_{\ell=1}^{M^{(a)}} \frac{q^{a+2}\,z^{2}-
\big(z^{(a)}_{\ell}\big)^{2}}{q^{a+1}\,z^2-q\,\big(z^{(a)}_{\ell}\big)^{2}}\
\frac{q^{a+2}\,\big(zz^{(a)}_{\ell}\big)^{2}-1}{q^{a+1}\,\big(zz^{(a)}_{\ell}\big)^{2}-q}
\ \prod_{\ell=1}^{M^{(a+1)}} \frac{q^{a}\,z^{2}-q\,
\big(z^{(a+1)}_{\ell}\big)^{2}}{q^{a+1}\,z^2-\big(z^{(a+1)}_{\ell}\big)^{2}}\
\frac{q^{a}\,\big(zz^{(a+1)}_{\ell}\big)^{2}-q}{q^{a+1}\,\big(zz^{(a+1)}_{\ell}\big)^{2}-1}
\nonu
&& a=0,\ldots,\enne-1 \label{dress2}
\end{eqnarray}
where we recall that $M^{(0)}=0$ and $M^{(\enne)}=0$. Finally by
requiring the vanishing of the residues at
$z=q^{-a/2}z_\ell^{(a)}$, we obtain the following set of Bethe
ansatz equations, for $a=1,\ldots,\enne$ and $k=1,\ldots,M^{(a)}$,
\begin{eqnarray}
  && \frac{g_{a}(z_{k}^{(a)}\,q^{-\frac{a}{2}}) }{
  g_{a+1}(z_{k}^{(a)} q^{-\frac{a}{2}})}\ \frac{\wh
  P_{a}\big(q^{\frac{a}{2}}/z_{k}^{(a)}\big)\ P_{a}(z_{k}^{(a)}\,q^{-\frac{a}{
  2}}) }{ \wh P_{a+1}\big(q^{\frac{a}{2}}/z_{k}^{(a)}\big)\
  P_{a+1}(z_{k}^{(a)}\,q^{-\frac{a}{2}})} \nonumber \\
  && \qquad\qquad
 =  -\prod_{\ell=1}^{M^{(a-1)}} \wh e_{-1}(z_{k}^{(a)};\
  z_{\ell}^{(a-1)}) \ \prod_{\ell=1}^{M^{(a)}} \wh
  e_{2}(z_{k}^{(a)};\ z_{\ell}^{(a)})\ \prod_{\ell=1}^{M^{(a+1)}}
  \wh e_{-1}(z_{k}^{(a)};\ z_{\ell}^{(a+1)})
  \qquad \label{BAE-open}
\end{eqnarray}
where we defined
\begin{equation}
  \wh e_{n}(z;w) = e_{n}(z/w)\ e_{n}(zw).
\end{equation}
Again, the right-hand side is linked to the $gl(\cN)$ Cartan matrix,
while the left-hand side is related to the chosen representations.
The choice of an open spin chain model is now determined by two
types of data:
\begin{enumerate}
\item The choice of the quantum spaces, i.e. of
 the Drinfel'd polynomials fixing the eigenvalues $P_{k}(z)$.
\item The choice of the boundary conditions, i.e. of $K(z)$ and of
$K^{+}(z)$, which fixes $g_{k}(z)$.
\end{enumerate}
Then, the spectrum of the transfer matrix is given by the solutions to
the BAE.

\subsubsection{Dressing and Cartan generators \label{sec:cartopen}}

As in the closed case, one can relate the dressing with the Cartan
generators eigenvalues.
 Starting from a transfer matrix eigenvector $v$, with eigenvalue
 \begin{equation}
b(z)\,v=\Lambda(z)\,v\ ,
\end{equation}
one can deduce, taking the limit $z\to\infty$, the eigenvalues for the
 Cartan generators of the finite dimensional symmetry algebra:
\begin{equation}
B_{jj}\,v = 2\,\Big( M^{(j-1)}-M^{(j)}+\sum_{n=1}^{N}
(a_{n}+\varpi_{j}^{(n)})\Big)\ v\ ,\
\forall\,j=1,\ldots,\enne\quad\mbox{with } M^{(0)}=0 
\mbox{ and }M^{(\enne)}=0\,.
\label{B-eigenvalue}
\end{equation}
This expression is valid whatever the representations on each site are.
Let us also remark that it is independent of the form of
the $K(z)$ and $K^+(z)$ matrices, i.e. of the boundary condition.

The upper bounds on the allowed values for the parameters $M^{(j)}$ is
deduced as in section \ref{M-borne} (with the same restrictions). We
get
\begin{equation}
\enne\, M_{j}\leq \sum_{k=1}^j k(\enne-j)\,n_{k} +\sum_{k=j+1}^{\enne-1}
j(\enne-k)\,n_{k}\ ,\ \forall j=1,\ldots,\enne-1\ .
\end{equation}
In the particular case of fundamental representations,
one recovers the condition
\begin{equation}
M^{(j)} \leq \frac{\enne-j}{\enne}\, N\ .
\end{equation}

\subsubsection{Example: open spin chain with one defect \label{example-open}}
As an example, we take the same spin chain as in the
closed spin chain case, section \ref{example-closed}, but now with open boundary
conditions given by a diagonal $K(z)$ matrix (\ref{eq:Kdiag}) and
$K^+(z)=M$.
The transfer matrix (based on $T_{0}(z)$ as in section
\ref{example-closed})
\begin{equation}
b(z)=tr_{0}\left(K^+_{0}(z) \,T_{0}(z)\, K_{0}(z)\, T_{0}^{-1}(\frac1z)\right)
\end{equation}
leads to a local Hamiltonian:
\begin{equation}
 \cH =\frac{1}{2\,(1-\xi^2)\,[\enne]_{q}}\,b'(1)=\sum_{\atopn{j=1}{j\neq
 p,p-1}}^{N-1} \cH_{j,j+1}+ \cH_{p-1,p,p+1}
 \end{equation}
with (we remind prime stands for $z$ derivative)
 \begin{eqnarray}
\cH_{j,j+1}&=&\frac{2}{q-q^{-1}}\,\cP_{j,j+1}\,R'_{j+1,j}\qmbox{,} 
j\neq p-1\,,\,p\,,\,N-1
 \\
\cH_{N-1,N}&=&\frac{1}{2(1-\xi^2)}\,K'_{N}(1)
\\[1.2ex]
\cH_{p-1,p,p+1}&=&L'_{p,p-1}(q^\theta)\,L^{-1}_{p,p-1}(q^\theta)
+\frac{1}{q-q^{-1}}\,L_{p,p-1}(q^\theta)\,
\cP_{p-1,p+1}\,R'_{p+1,p-1}\,L^{-1}_{p,p-1}(q^\theta)\,\qquad
 \qquad\\[1.2ex]
  R'_{ab}&=&2\big(\II-P_{ab}^q\big)+(q+q^{-1})\,\cP_{ab}\,,\quad
  \forall\ a\neq b.
 \end{eqnarray}
 To obtain these results, we have used 
 $$
 K(1)=(1-\xi^2)\,\II \qmbox{and} tr(M)=[\enne]_{q}\equiv
 \frac{q^{\enne}-q^{-\enne}}{q-q^{-1}}
 $$
 The Hamiltonian eigenvalues are of the form 
 $\wt E=\frac{1}{2\,(1-\xi^2)\,[\enne]_{q}}\,\Lambda'(1)$, where
 $\Lambda(z)$ is now given in (\ref{eq:Lambda-open}), with dressings
 (\ref{dress2}), BAEs (\ref{BAE-open}) and 
 \begin{eqnarray*}
 \pi_{a}(z) &=&
 zq^{\varpi_{a}+\theta}-z^{-1}q^{-\varpi_{a}-\theta}\qquad
\qquad\quad \forall a\\[1.2ex]
 P_{a}(z) &=& \left\{\begin{array}{ll}
 (zq-z^{-1}q^{-1})^{N-1}\,\pi_{1}(z)
 &\qmbox{if} a=1 \\[1.2ex]
  (z-z^{-1})^{N-1}\, \pi_{a}(z)
 &\qmbox{if} a\neq1
 \end{array}\right.\\[1.2ex]
\wh P_{a}(z) &=& \left\{\begin{array}{ll} P_{1}(z)^{-1}
 &\qmbox{if} a=1 \\[1.2ex]
 \displaystyle
 \left(\frac{z-z^{-1}}{(zq-z^{-1}q^{-1})(zq^{-1}-z^{-1}q)}\right)^{N-1}
 \left(\prod_{j=1}^{a-1}\frac{\pi_{j}(zq^{-j})}{\pi_{j}(zq^{-j+1})}\right)\,
\frac{1}{\pi_{a}(zq^{-a+1})} &\qmbox{if} a\neq1
 \end{array}\right.
 \end{eqnarray*}
Using these expressions, one gets
\begin{eqnarray}
    \wt E &=& \wt E_{0}+2\,(q^2-1)\,\sum_{\ell=1}^{M^{(1)}}
\frac{\big(z_{\ell}^{(1)}\big)^2}{ \left(1-q\,\big(z_{\ell}^{(1)}\big)^2\right)
\left(q-\big(z_{\ell}^{(1)}\big)^2\right)}\\[1.2ex]
\wt E_{0} &=& (N-2)\,
\frac{q+q^{-1}}{q-q^{-1}}+\frac{2\,q^\enne}{q^\enne-q^{-\enne}}
+\frac{q^{\varpi_{1}+\theta}+q^{-\varpi_{1}-\theta}}
{q^{\varpi_{1}+\theta}-q^{-\varpi_{1}-\theta}}+\frac{1}{1-\xi^2}
\end{eqnarray}
where $\wt E_{0}$ is the energy of the open spin chain pseudo-vacuum.
\section{Quantum twisted Yangians and SNP spin chains}

As already mentioned, the construction of the reflection algebra
(hence of "soliton preserving" open spin chains) is available for any
quantum group. This property is based on the (formal) existence for
any algebra of the inversion antimorphism ${\cal L}(z)\to {\cal
L}^{-1}(z)$. However, there are some algebras, such as the
Yangian of $gl(\enne)$, or the quantum algebra  $\cU_{q}(\wh{gl}({\cal
N}))$, for which another
antimorphism can be constructed, leading to other (new) coideal
subalgebras, which themselves allow one to
construct (new) open spin chains. These spin chains have
different boundaries, known as soliton non-preserving boundaries.
These physical considerations will be developed below.

\subsection{The quantum twisted Yangian\label{sect:qYtw}}

We first focus on the mathematical construction, which will infer
the monodromy matrix of the aforementioned spin chains. The algebra
we construct is the quantum twisted Yangian, as it has been
introduced in \cite{qYangtw}.

\subsubsection{Definition}

We start with the transposition in the auxiliary space:
\begin{equation}
 ^{t_{0}}\ :\ \cL_{01}(z)=\sum_{a,b=1}^\enne E_{ab}\otimes L_{ab}(z)\ \mapsto\
 \cL_{01}^{t_{0}}(z)=\sum_{a,b=1}^\enne E_{ab}\otimes L_{ba}(z)
 \end{equation}
which is an antimorphism of ${\wh\cU_{q}}$. Mimicking the construction of the
quantum reflection algebra, we introduce\footnote{One could also
take $\cS^\pm(z)=\cL^\pm(z)G\cL^\pm(z^{-1})^t$ which obey the same
exchange relations: as for quantum reflection algebra, it leads to
an equivalent construction for spin chains, up to analytical
properties of $\cS(z)$.}:
\begin{eqnarray}
  \cS(z) &=& \cL^+(zq^{-\frac{c}{2}})\,G\,\cL^-(z^{-1}q^{\frac{c}{2}})^t
 \quad\mbox{where }\label{defS}
 \left\{\begin{array}{l}
G =\II\quad o(\enne)\mbox{ case, }\\[1.2ex]
\displaystyle G = \sum_{k=1}^n\Big( q\,E_{2k-1,2k}-E_{2k,2k-1}\Big)
\quad sp(2n)\mbox{ case }.
\end{array}\right.
\end{eqnarray}

It is easy to show that $\cS(z)$ defines a subalgebra of
$\cU_{q}(\wh{gl}(\enne))$, called quantum twisted Yangian
\cite{qYangtw}, with exchange relations (when $c=0$)
\begin{equation}
  R_{12}(\frac{z}{w})\,\cS_{1}(z)\,
  R_{12}^{t_{1}}(\frac{1}{zw})\,\cS_{2}(w) =
  \cS_{2}(w)\,R_{12}^{t_{1}}(\frac{1}{zw})\,
  \cS_{1}(z)\,R_{12}(\frac{z}{w})\,,\label{rsrs}
\end{equation}
together with the conditions
\begin{eqnarray}
o(\enne)\mbox{ case:} &&\left\{\begin{array}{l}
\cS_{ii}(0) = 1\,,\ \forall\ i \\[1.2ex]
\cS_{ij}(0) = 0\,, \ \forall\ i>j
\end{array}\right.
\\[1.2ex]
sp(\enne)\mbox{ case:} &&\left\{\begin{array}{l}
\cS_{2k,2k}(0)\,\cS_{2k-1,2k-1}(0)-q^{2}\,\cS_{2k,2k-1}(0)\,\cS_{2k-1,2k}(0) =
q^{3}\,,\ \forall\ k\\[1.2ex]
\cS_{ij}(0) = 0 \,,\ \forall\ i>j\mbox{ and } i\neq j'
\end{array}\right.
\end{eqnarray}
We have introduced the map
\begin{equation}
(2k)'=2k-1\,,\ (2k-1)'=2k\quad\mbox{i.e.}\quad
j'=2\,\left[\frac{j+1}{2}\right]+\left[\frac{j-1}{2}\right]-\left[\frac{j}{2}\right]
\,.
\end{equation}
Note that the above calculation  uses the equality (valid for both
values of $G$):
\begin{eqnarray}
&& R_{12}(z)\,G_{1}\,R_{12}^{t_{1}}(w)\,G_{2} =
G_{2}\,R_{12}^{t_{1}}(w)\,G_{1}\,R_{12}(z),
\end{eqnarray}
which have to be compared with the relation (\ref{supR}) used in the
case of quantum reflection algebras.

Existence of constant solutions to the relation (\ref{rsrs}) is
ensured by the CP-invariance (\ref{CP-inv2}), which proves that any
invertible diagonal matrix is a solution.

Then, the construction follows the lines of section \ref{sect:qrefl},
see \cite{twisty}. For instance one has
\begin{eqnarray}
\cS(z) &=& \cL^+(zq^{-\frac{c}{2}})\,G(z)\,\cL^-(z^{-1}q^{\frac{c}{2}})^t
\label{gensol:S}\\
  \Delta({\cal S}(z)) &=&  {\cal
  L}_{01}(zq^{-\frac{c_{1}}{2}})\,
  {\cS}_{02}(zq^{-c_{1}})\, {\cal
  L}^{t}_{01}(q^{\frac{c_{1}}{2}}z^{-1})
\end{eqnarray}
where $G(z)$ is a solution of the reflection equation (\ref{rsrs})
such that $G(0)=G$ (up to invariance of the exchange relation).

Here, keeping in mind the spin chain interpretation, we
will use an equivalent definition of the quantum twisted Yangian. The
exchange relations we choose have the following form
\begin{equation}
  R_{12}(\frac{z}{w})\ \wt\cS_{1}(z)\ \bar
  R_{21}(zw)\ \wt\cS_{2}(w)=
  \wt\cS_{2}(w)\ \bar R_{12}(zw)\
  \wt\cS_{1}(z)\ R_{21}(\frac{z}{w}),
  \label{newTwY}
\end{equation}
where the $\bar R$ matrices are defined in appendix \ref{sect:Rbar}.
The relation between the two definitions is given by
\begin{equation}
  \cS(z)=\wt\cS(z\rho^{-\frac{1}{2}})V^t.
  \label{toto}
\end{equation}
The algebra generated by $\wt\cS(z)$ can be seen as a subalgebra of
${\wh\cU_{q}}$ through the construction ($c=0$)
\begin{equation}
  \wt\cS(z)=\cL^{+}(z\rho^{\frac{1}{2}})\,\wt K(z)\,
  V^t\,\cL^-(z^{-1}\rho^{-\frac{1}{2}})^t\,V^t\,,
  \label{defKtilde}
\end{equation}
where $\wt K(z)$ is any $\CC$-valued matricial solution of the
relation (\ref{newTwY}). Of course, the construction (\ref{defKtilde}) can
be deduced from (\ref{gensol:S}) using the relation (\ref{toto}).
The existence of solutions is ensured by the relation (\ref{toto}), which
proves that $W\,V^t$ (where $W$ is diagonal) is indeed a solution. \\
In this formalism, the coproduct takes the form
\begin{equation}
  \Delta (\wt \cS(z))={\cal L}_{01}(z\rho^{-\frac{1}{2}}q^{-\frac{c_{1}}{2}})\,
  {\wt \cS}_{02}(zq^{-c_{1}})\, V_{0}^t{\cal
  L}^{t}_{01}(z^{-1}\rho^{-\frac{1}{2}}q^{\frac{c_{1}}{2}})V_{0}^t\,.
\end{equation}
The  universal monodromy matrix will then be ($c$=0)
\begin{eqnarray}
  \Delta^{(N)} (\wt \cS(z))=\cL_{01}(z\rho^{\frac{1}{2}})\cdots
  \cL_{0N}(z\rho^{\frac{1}{2}})\,\wt K_{0}(z)\,
  V_{0}^t\,\cL^{t_{0}}_{0,N}(z^{-1}\rho^{-\frac{1}{2}})\cdots
  \cL^{t_{0}}_{01}(z^{-1}\rho^{-\frac{1}{2}})\,V_{0}^t\,.
\end{eqnarray}
The  corresponding transfer matrix is defined as
\begin{equation}
  \wt s(z) = tr_{0}\ \Big \{\wt K_{0}^+(z)\,\Delta^{(N)}(
  \wt\cS_{0}(z))\Big \}
  \label{transferSNP}
\end{equation}
where  $\wt K^+(z)$ satisfy the
dual reflection equation
\begin{eqnarray}
  && R_{12}(\frac{w}{z})\ \wt{K}_{1}^{+}(z)^t\
  M_{1}^{-1}\ \bar R_{21}(\rho^{-2}z^{-1}w^{-1})\ M_{1}\ \wt
  K_{2}^{+}(w)^t= \nonumber \\
  && \wt K^{+}_{2}(w)^t\ M_{1}\ \bar
  R_{12}(\rho^{-2}z^{-1}w^{-1})\ M_{1}^{-1}\ \wt
  K_{1}^{+}(z)^t\ R_{21}(\frac{w}{z}), \label{snd}
\end{eqnarray}
Solutions to the two equations (\ref{newTwY}) and (\ref{snd}) are
related through $\wt K^{+}(z)=M\,\wt K^t(z^{-1}\rho^{-\frac{1}{2}})$,
so that $\wt K^+(z)=V^t$ is also a solution to (\ref{snd}).\\
Again, it can be proved using only the
 equations (\ref{newTwY}), (\ref{snd}),  the unitarity of
the $R$ matrix, and arguments similar to the ones given in \cite{sklyanin}
(for the SP case), that
\begin{equation}
  \Big [ \wt s(z),\ \wt s(w) \Big ]=0\;.
\end{equation}
It will ensure that the open spin chain derived from (\ref{transferSNP}) is also
integrable.

\subsubsection{Finite dimensional subalgebras of the quantum twisted
Yangian}
Starting from the exchange relation (\ref{newTwY}), one gets
\begin{equation}
  R_{12}\ \wt\cS_{1}(z)\ \bar R_{21}\ \wt\cS_{2}(0)=
  \wt\cS_{2}(0)\ \bar R_{12}\ \wt\cS_{1}(z)\ R_{21},
  \label{SzS0}
\end{equation}
where $\bar R_{12}=V_{1}\,R_{12}^{t_{2}}\,V_{1}$. It proves that
$\wt\cS(0)$ form a subalgebra of the twisted Yangian, and that
$\wt\cS(z)$ is a representation of it. When $\wt K(z)=G\,V^t$, this subalgebra
(up to the change of basis due to (\ref{toto}))
is nothing but the the twisted quantum algebra $\cU^{tw}_{q}(\fg)$
introduced in \cite{n:ms},
with $\fg=o(\enne)$ if $\theta_{0}=1$ or $\fg=sp(\enne)$ if
$\theta_{0}=-1$. It corresponds to a deformation of the classical
algebra $\cU(\fg)$, different from the quantum algebra $\cU_{q}(\fg)$.
In particular, it has no (known) proper Hopf structure, but is rather
a Hopf coideal of $\cU_{q}(gl(\enne))$.
Let us remark that when $\wt K$ is an antidiagonal matrix,
$\wt S(0)$ is equivalent\footnote{In fact, it is
$\cS(0)=\wt \cS(0)\,V^t$ which is triangular when $G$ is diagonal. $\wt
\cS(0)$ is triangular with respect to the second diagonal.} to a lower
triangular ($so(\enne)$ case) or block triangular ($sp(2n)$ case)
matrix,
in accordance with the dimension of $\cU^{tw}_{q}(\fg)$.

\null

Defining $\wh \cS(z)=\wt \cS(z^{-1})^{-1}$, one gets
\begin{eqnarray}
&& \bar R_{12}(\frac{z}{w})\ \wt\cS_{1}(z)\ R_{21}(zw)\ \wh\cS_{2}(w)=
  \wh\cS_{2}(w)\ R_{12}(zw)\ \wt\cS_{1}(z)\ \bar
  R_{21}(\frac{z}{w}),
    \label{Stild-Shat}\\
&& R_{12}(\frac{z}{w})\ \wh\cS_{1}(z)\ \bar R_{21}(zw)\ \wh\cS_{2}(w)=
\wh\cS_{2}(w)\ \bar R_{12}(zw)\ \wh\cS_{1}(z)\ R_{21}(\frac{z}{w}),
\end{eqnarray}
which shows that $\wh \cS(z)$ obeys the same relation as $\cS(z)$.
One can, as in section \ref{sect:finiteSalg}, consider $\wh \cS(\infty)$
(if $\wt K(z)$ is a constant antidiagonal matrix), or some
regularisation of it (when $\wt K(z)$ depends on $z$). Its exchange
relations are given by
\begin{eqnarray}
&& \bar R_{12}\ \wt\cS_{1}(z)\ R_{21}\ \wh\cS_{2}(\infty)=
  \wh\cS_{2}(\infty)\ R_{12}\ \wt\cS_{1}(z)\ \bar R_{21},
\label{Stild-Shat0}\\
&& R_{21}^{-1}\ \wh\cS_{1}(z)\ \wt R_{21}\ \wh\cS_{2}(\infty)=
\wh\cS_{2}(\infty)\ \wt R_{12}\ \wh\cS_{1}(z)\ R_{12}^{-1},\\
&& \wt R_{12}=V_{1}\,\big(R_{21}^{-1}\big)^{t_{2}}\,V_{1}
=V^t_{1}\,\big(R_{12}^{t_{1}}\big)^{-1}\,V^t_{1}
\end{eqnarray}
\begin{lemma}
 The generators $\wt\cS(0)$ and $\wh\cS(\infty)$ satisfy
$ \wt\cS(0)\,\wh\cS(\infty)=\wh\cS(\infty)\,\wt\cS(0)=x_{0}\,\II$
 where $x_{0}$ is some constant.\\
 In particular, one has
  \begin{equation}
{[\wt\cS(0)\,,\,\wh\cS(\infty)]}=0,
 \end{equation}
so
 that either $\wh\cS(\infty)$ is the inverse of $ \wt\cS(0)$, or
 generates a subalgebra which commutes with the one generated by $ \wt\cS(0)$.
 \end{lemma}
 \underline{Proof:} One starts from the relation (\ref{Stild-Shat})
 and consider the limits $z\to0$ then $w\to\infty$ on the one hand,
 and $w\to\infty$ then $z\to0$ on the other hand. It leads to the
 relations
\begin{eqnarray}
&& \bar R_{12}\ \wt\cS_{1}(0)\ R_{12}^{-1}\ \wh\cS_{2}(\infty)=
  \wh\cS_{2}(\infty)\ R_{21}^{-1}\ \wt\cS_{1}(0)\ \bar R_{21},\\
&& \bar R_{12}\ \wt\cS_{1}(0)\ R_{21}\ \wh\cS_{2}(\infty)=
  \wh\cS_{2}(\infty)\ R_{12}\ \wt\cS_{1}(0)\ \bar R_{21}.
\end{eqnarray}
Using relation (\ref{eq:R12-R21inv}), it leads to
\begin{equation}
  \bar R_{12}\ \wt\cS_{1}(0)\ \wh\cS_{1}(\infty)=
  \wh\cS_{2}(\infty)\ \wt\cS_{2}(0)\ \bar R_{12},
\end{equation}
which gives after some manipulations
\begin{equation}
  V_{1}\ \wt\cS_{1}(0)\ \wh\cS_{1}(\infty)\ V_{1}=
 \left( \wh\cS_{2}(\infty)\ \wt\cS_{2}(0)\right)^t\ ,
\end{equation}
whose only solution is the one given in the lemma.\finproof
Remark that this lemma is the counterpart of lemma \ref{lem:Bo+Binf}.

We leave to the
interested reader the complete study of the subalgebra spanned by 
 $\wt\cS(0)$ and $\wh\cS(\infty)$. It is related to twisted Yangians
 $\cU_{q}^{tw}(\fh)$, where $\fh$ is a subalgebra of $o(\enne)$ or
 $sp(\enne)$ which depends on the chosen
  $\wt K(z)$ matrix. We classify these latter in section
 \ref{sect:SNP}.

\subsubsection{Symmetry of the universal transfer matrix}
To study the symmetry of the transfer matrix, one should proceed
as in the SP case. However, there is a crucial difference between the
relations (\ref{CR:B0}) and (\ref{SzS0}): in the first case, each side of the
equality involves an $R$ matrix and essentially its inverse, while it
is essentially its transpose in the second case. Then, it is not
possible to get an equivalent of lemma \ref{lem:tr1} in the SNP
context. Indeed, starting from (\ref{rsrs}), it is easy to conclude
that
\begin{equation}
 \left[\,tr_{1}\big(R_{12}\,\cS_{1}(z)\, R_{12}^{t_{1}}\big)\ ,\
 \cS_{2}(0)\,\right] =0\,,
\end{equation}
but $tr_{1}\big(R_{12}\,\cS_{1}(z)\, R_{12}^{t_{1}}\big)$ is not
easily related to $\wt s(z)$, so that one cannot obtain, in 
this way, a symmetry of the transfer matrix. Similar conclusion is
obtained when starting from (\ref{SzS0}) or (\ref{Stild-Shat0}). As a
consequence, it looks as if the SNP spin chain does not possess any symmetry.
In fact, the matrices $\cS(z)$, $\wt\cS(z)$ and $\wh\cS(z)$ are not
written in a Cartan-Weyl basis, so that even the Cartan generators
(which should commute with $\wt s(z)$) are difficult to identify in
this presentation. 

\subsection{Soliton non-preserving boundaries\label{sect:SNP}}

In this section we shall deal with the so-called soliton non-preserving
boundary conditions (SNP) \cite{doikou1, selene, twisty}, which physically
describe the reflection of a soliton to an anti-soliton. The corresponding
equation describing such boundary conditions is given \cite{gand,dema} 
by (\ref{newTwY}).
The main aim now is to derive matricial solutions of this equation.\\
It is convenient to consider the matrix
$G(z)=\wt{K}(z\rho^{-\frac{1}{2}})\,V^t$ instead of $\wt K(z)$, which obeys
the reflection equation (\ref{rsrs}).
\begin{lemma}
\label{lemma:DKD}
If $G(z)$ is a solution of the reflection equation (\ref{rsrs}), then
$DG(z)D$ is also a solution of (\ref{rsrs}), where $D$ is any constant
diagonal matrix.
\end{lemma}
\underline{Proof:} one multiplies (\ref{rsrs}) on the right and on the left
by $D_{1}\,D_{2}$ and uses the property (\ref{CP-inv2}) of 
the matrix $R_{12}(z)$.
\finproof
This property was already noticed in \cite{gand}.
\begin{proposition}
The invertible solutions to the reflection equation (\ref{newTwY}) must be, up
to a normalization factor, of the form  
\begin{eqnarray}
\wt K(z)=DG(z\rho^{\frac{1}{2}})D\,V^t
\label{eq:ktilde-k}
\end{eqnarray}
where $D$ is any constant invertible diagonal matrix and $G(z)$ is one of
the following matrices:
\begin{eqnarray}
  &*& G(z) \text{ is the unit matrix} 
  \label{solSNP:un} \\
  &*& G(z) = \frac{q^2-z^4}{q+1} \sum_{i=1}^\enne E_{ii} - z^2(z^2\pm q)
  \sum_{i<j} E_{ij} + (q\pm z^2) \sum_{i>j} E_{ij} 
  \label{solSNP:z} \\
  &*& G(z) = \sum_{i=1}^{\enne/2} (qE_{2i-1,2i} - E_{2i,2i-1}) = G \qquad
  \text{when $\enne$ is even}
  \label{solSNP:G} \\
  &*& G(z) = z^2 E_{1\enne} - qz^{-2} E_{\enne 1} + \sum_{i=1}^{(\enne-2)/2}
  (qE_{2i,2i+1} - E_{2i+1,2i}) \qquad \text{when $\enne$ is even}
  \label{solSNP:AG} 
\end{eqnarray}
In the case of $gl(4)$, there is one additional solution:
\begin{equation}
G(z) = E_{13} \pm z^{-2} E_{31} + E_{24} \pm z^{-2} E_{42} 
\label{eq:solsup}
\end{equation}
Note that the solution (\ref{solSNP:AG}) takes an antidiagonal form in the
$gl(4)$ case: 
\begin{equation}
G(z) =z^2\,E_{14}+q\,E_{23}-\,E_{32}-qz^{-2}\,E_{41}
\label{eq:solsupdiag}
\end{equation}
\end{proposition}
\underline{Proof:} The matrix $\displaystyle G(z) =
\wt{K}(z\rho^{-\frac{1}{2}})V^t = \sum_{i,j=1}^\enne G_{ij}(z)\,E_{ij}$
obeys the reflection equation (\ref{rsrs}). \\
The projection of (\ref{rsrs}) on $E_{ii} \otimes E_{jj}$ with $i<j$ reads
\begin{eqnarray}
&& G_{ij}(w) \left(\frac{q^2-z^2w^2}{qw^2(z^2-w^2)} \, G_{ji}(z) +
\frac{1}{z^2w^2} \, G_{ij}(z) \right) = G_{ji}(w) \left(
\frac{q^2-z^2w^2}{qz^2(z^2-w^2)}\, G_{ij}(z) + G_{ji}(z) \right) \qquad
\label{eq:EiiEjjbis}
\end{eqnarray}
which shows that $G_{ij}(z) = 0 \Leftrightarrow G_{ji}(z) = 0$. Moreover,
the relation (\ref{eq:EiiEjjbis}) implies that the non-vanishing
off-diagonal elements $G_{ij}(z)$ and $G_{ji}(z)$ satisfy ($\alpha_{ij}$
and $\beta_{ij}$ are free parameters)
\begin{equation}
G_{ji}(z) = F_{ij}(z)\,G_{ij}(z) 
\qquad \text{with} \qquad F_{ij}(z) =
z^{-2}\;\frac{\alpha_{ij}z^2-q\beta_{ij}} {\beta_{ij}z^2-q\alpha_{ij}}
\qquad ,\ i<j
\label{eq:fij}
\end{equation}
Then projecting (\ref{rsrs}) on the $E_{ij} \otimes E_{ij}$ component (with
$i \ne j$), one gets
\begin{equation}
G_{ii}(z)\,G_{jj}(w) = G_{ii}(w)\,G_{jj}(z)
\label{eq:EijEij}
\end{equation}
which implies that $G_{ii}(z)$ can be taken as constant numbers (note that
it is still true if there is only one non-vanishing diagonal element since
the matrix $G(z)$ is defined up to an overall normalization factor).

\medskip

$\blacktriangleright$ 
Let us first assume that all diagonal elements $G_{ii}$ are non-zero. Then,
the $E_{ii} \otimes E_{ij}$ and $E_{ii} \otimes E_{ji}$ components of
(\ref{rsrs}) lead to (with $i<j$)
\begin{eqnarray}
&& \frac{q^2-w^4}{w^2} \, G_{ij}(w) - \frac{q^2-z^2w^2}{z^2} \, G_{ij}(z) -
q(z^2-w^2) G_{ji}(z) = 0 \\
&& \frac{q^2-w^4}{w^2} \, G_{ji}(w) - \frac{q^2-z^2w^2}{w^2} \, G_{ji}(z) -
q\,\frac{z^2-w^2}{z^2w^2} \, G_{ij}(z) = 0
\end{eqnarray}
from which it follows 
\begin{equation}
G_{ij}(z) = qz^2\,\frac{\beta_{ij} z^2 - q\alpha_{ij}}{z^4-q^2} \quad
\mbox{and}\quad G_{ji}(z) = q\,\frac{\alpha_{ji} z^2 - q\beta_{ji}}
{z^4-q^2}\qquad,\ i < j
\label{eq:Kij}
\end{equation}
where $\alpha_{ji} = \alpha_{ij}$ and $\beta_{ji} = \beta_{ij}$. \\
Now, one projects (\ref{rsrs}) on the components $E_{ij} \otimes E_{kk}$
($i \ne j \ne k$):
\begin{eqnarray}
 && (1-z^2w^2) \Big( \big( z^2\,\delta_{i<k} +
w^2\,\delta_{i>k} \big) G_{kj}(z) G_{ik}(w) - \big( w^2\,\delta_{j<k} +
z^2\,\delta_{j>k} \big) G_{ik}(z) G_{kj}(w) \Big) \qquad\nonumber \\
&& + (z^2-w^2) \Big( \big( \delta_{j<k} +
z^2w^2\,\delta_{j>k} \big) G_{ik}(z) G_{jk}(w) - \big( z^2w^2\,\delta_{i<k}
+ \delta_{i>k} \big) G_{kj}(z) G_{ki}(w) \Big) = 0\qquad\qquad
\label{eq:EijEkk}
\end{eqnarray}
and on the components $E_{ij} \otimes E_{kl}$ ($i,j,k,l$ all distinct):
\begin{eqnarray}
&& (1-z^2w^2) \Big( \big( z^2\,\delta_{i<k} + w^2\,\delta_{i>k} \big)
G_{kj}(z) G_{il}(w) - \big( w^2\,\delta_{j<l} + z^2\,\delta_{j>l} \big)
G_{il}(z) G_{kj}(w) \Big) \nonumber \\
&& + (z^2-w^2) \Big( \big( \delta_{j<k} + z^2w^2\,\delta_{j>k} \big)
G_{ik}(z) G_{jl}(w) - \big( z^2w^2\,\delta_{i<l} + \delta_{i>l} \big)
G_{lj}(z) G_{ki}(w) \Big) \nonumber \\
&& + (q-q^{-1})z^2w^2 \Big( \big( z^2\,\delta_{i<(j,k)} +
w^2\,\delta_{k<i<j} + w^{-2}\,\delta_{k>i>j} +z^{-2}\,\delta_{i>(j,k)}
\big) G_{ki}(z) G_{jl}(w) \qquad\nonumber \\
&&\hspace{7.1em}
- \big( z^2\,\delta_{(i,l)<j} + w^2\,\delta_{i<j<l} +
w^{-2}\,\delta_{i>j>l} +z^{-2}\,\delta_{(i,l)>j} \big) G_{jl}(z) G_{ki}(w)
\Big) = 0\qquad\quad
\label{eq:EijEkl}
\end{eqnarray}
where $\delta_{m<n} = 1$ if $m<n$ and zero otherwise (and similarly 
$\delta_{(m,p)<n} = 1$ if $m<n$ and $p<n$ and zero otherwise, and so on). \\
Inserting expressions (\ref{eq:Kij}) into (\ref{eq:EijEkk}) and
(\ref{eq:EijEkl}), one gets
\begin{equation}
\alpha_{ik}\,\beta_{jl} = \beta_{ik}\,\alpha_{jl} \,, \;\; i<(j,k)<l \quad
\mbox{and} \quad \alpha_{ik}\,\alpha_{jl} = \beta_{ik}\,\beta_{jl} \,, \;\;
(i,j)<(k,l)
\end{equation}
whose solution is given by $\alpha_{ij} = \epsilon\,\beta_{ij}$ for all
pair $(i,j)$ of indices ($\epsilon=\pm 1$), with the consistency condition
\begin{equation}
\beta_{ij}\,\beta_{kl} = \beta_{ik}\,\beta_{jl} = \beta_{il}\,\beta_{jk}
\qquad (i<j<k<l)
\label{eq:cijckl}
\end{equation}
The projection on the components $E_{ij} \otimes E_{ik}$:
\begin{eqnarray}
&& \frac{(1-z^2w^2)(qz^2+w^2)}{q+1} \, G_{ij}(z) G_{ik}(w) - (1-z^2w^2)
\big( w^2\,\delta_{j<k} + z^2\,\delta_{j>k} \big) G_{ij} (w) G_{ik}(z)
\nonumber \\
&& \; + \; \Big( (qz^2-q^{-1}w^2) \big( z^2w^2\,\delta_{i<j} + \delta_{i>j}
\big) G_{jk}(w) - (z^2-w^2) \big( z^2w^2\,\delta_{i<k} + \delta_{i>k} \big)
G_{kj}(z) \Big. \nonumber \\
&& \; - \; \Big. (q-q^{-1}) z^2w^2 \big( z^2\,\delta_{(i,k)<j} +
w^2\,\delta_{i<j<k} + w^{-2}\,\delta_{i>j>k} + z^{-2}\,\delta_{(i,k)>j}
\big) G_{jk}(z) \Big) G_{ii} = 0
\label{eq:EijEik}
\end{eqnarray}
then leads to the conditions (with $i,j,k$ distinct indices)
\begin{equation}
(q+1)\,G_{ii}\,\beta_{jk} = q\,\beta_{ij}\,\beta_{ik} 
\label{eq:kiibeta}
\end{equation}
Since $G_{ii} \ne 0$ for all $i$, equation (\ref{eq:kiibeta}) implies that
the coefficients $\beta_{ij}$ (and therefore $\alpha_{ij}$) are either all
zero, or all non-zero. In the first case, $G(z)$ is a constant diagonal
matrix (which can be brought to the unit matrix due to lemma
\ref{lemma:DKD}). In the second case, equation (\ref{eq:kiibeta}) is solved
as
\begin{equation}
\beta_{ij}^2 = \frac{(q+1)^2}{q^2} \, G_{ii}\,G_{jj}
\end{equation}
Thanks to lemma \ref{lemma:DKD}, one can bring the matrix $G(z)$ in the
form (\ref{solSNP:z}). Finally, one can check that all remaining equations
obtained from (\ref{rsrs}) are satisfied. \\
These two cases corespond to the matrices (\ref{solSNP:un}) and
(\ref{solSNP:z}) exhibited in the classification (up to multiplication by a
function, so that $G(z)$ is analytical).

\medskip

$\blacktriangleright$ 
We now consider the case where there is at least one diagonal element
$G_{ii}$ which is zero. The projections of (\ref{rsrs}) on $E_{ij} \otimes
E_{ik}$ and $E_{ji} \otimes E_{ki}$ reduce to
\begin{eqnarray}
\label{eq:EijEikred}
&& (qz^2+w^2) G_{ij}(z) G_{ik}(w) - (q+1)\big( w^2\,\delta_{j<k} + z^2 \,
\delta_{j>k} \big) G_{ij}(w) G_{ik}(z) = 0 \\
\label{eq:EjiEkired}
&& (qz^2+w^2) G_{ji}(z) G_{ki}(w) - (q+1)\big( z^2\,\delta_{j<k} + w^2 \,
\delta_{j>k} \big) G_{ji}(w) G_{ki}(z) = 0
\end{eqnarray}
which imply $G_{ij}(z)\,G_{ik}(w) = 0$ and $G_{ji}(z)\,G_{ki}(w) = 0$
for each triple $(i,j,k)$ of distinct indices. Since $G_{ij}(z) = 0
\Leftrightarrow G_{ji}(z) = 0$, it follows that the reflection matrix
$G(z)$ has at most one non-zero element in $i$-th row and in $i$-th column.
We assume that $G_{ij}(z) \ne 0$ for some $j \ne i$ (otherwise the matrix
$G(z)$ is not invertible). \\
Considering now the projections on $E_{in} \otimes E_{mj}$ and on $E_{im}
\otimes E_{mj}$, one gets for $m,n \ne i,j$
\begin{equation}
\big( z^2\,\delta_{i<m} + w^2\,\delta_{i>m} \big) G_{mn}(z) G_{ij}(w) -
\big( z^2\,\delta_{j<n} + w^2\,\delta_{j>n} \big) G_{ij}(z) G_{mn}(w) = 0
\label{eq:EijEklbis}
\end{equation}
{from} which one can deduce that there are complex numbers $\mu_{mn}$ 
such that
\begin{equation}
  G_{mn}(z) =
  \begin{cases}
    \mu_{mn} \, G_{ij}(z) & \text{for $m<i$ and $n<j$ or $m>i$ and $n>j$}
    \\
    \mu_{mn} \, z^2 \, G_{ij}(z) & \text{for $m<i$ and $n>j$} \\
    \mu_{mn} \, z^{-2} \, G_{ij}(z) & \text{for $m>i$ and $n<j$}
  \end{cases}
  \label{eq:kmn}
\end{equation}
Now, the $E_{ji} \otimes E_{jk}$ and $E_{ij} \otimes E_{kj}$ components of
(\ref{rsrs}) ($i,j,k$ distinct) lead to equations similar to
(\ref{eq:EijEikred}) and (\ref{eq:EjiEkired}), relating the entries
$G_{ij}$ and $G_{kj}$ on the one hand and $G_{ji}$ and $G_{jk}$ on the
other hand. They imply that $G_{jk}(z) = G_{kj}(z) = 0$ for $k \ne i,j$.
Hence, the matrix $G(z)$ exhibits the following shape (taking e.g. $i<j$):
\begin{equation}
G(z) = \left(
\begin{array}{ccccc}
  (M_{11}) & 0 & (M_{12}) & 0 & (M_{13}) \\
  0 & 0 & 0 & G_{ij}(z) & 0 \\
  (M_{21}) & 0 & (M_{22}) & 0 & (M_{23}) \\
  0 & G_{ji}(z) & 0 & G_{jj}(z) & 0 \\
  (M_{31}) & 0 & (M_{32}) & 0 & (M_{33}) \\
\end{array}
\right)
\label{matshape}
\end{equation}
where $(M_{mn})$ represent block submatrices whose entries are given by
(\ref{eq:kmn}). \\
Exchanging the r\^{o}les of the indices $i$ and $j$, and taking into
account the relations (\ref{eq:kmn}), one obtains the following necessary
conditions for the matrix $G(z)$ to be invertible: 
\begin{itemize}
  \item[(i)]
  If $|j-i|>1$ with $i \ne 1$ and $j \ne \enne$, the block submatrices
  $(M_{13})$, $(M_{31})$ and $(M_{mm})$ are identically zero and the
  function $F_{ij}(z)$ must be equal to $\pm z^{-2}$ $(i<j)$, see eq.
  (\ref{eq:fij}).
  \item[(ii)]
  If $|j-i|>1$, with $i = 1$ or $j = \enne$ (exclusively), the conditions
  are similar to the case (i), but the block submatrices $(M_{1m})$ for
  $i=1$ or $(M_{m3})$ for $j=\enne$ are not present.
  \item[(iii)]
  If $i = 1$ and $j = \enne$, only the block submatrix $(M_{22})$
  survives and the function $F_{1\enne}(z)$ is equal to $-qz^{-4}$, see eq.
  (\ref{eq:fij}).
  \item[(iv)]
  If $j=i+1$, the block submatrices $(M_{2m})$ and $(M_{m2})$ are not
  present and the function $F_{i,i+1}(z)$ is equal to $-q^{-1}$, see eq.
  (\ref{eq:fij}).
\end{itemize}
$\bullet$ \textit{We first treat the case (i)}, where $|j-i|>1$ with $i,j \ne
1,\enne$. 
Due to the vanishing of $(M_{mm})$, the diagonal elements $G_{kk}$ are 
zero for
all $k \ne j$. Hence, from the above arguments, the matrix $G(z)$ has
exactly one non-zero off diagonal element in each row and in each column to
be invertible. One can easily prove by recursion that there is no such
matrix when $\enne$ is odd. Since we ask for invertible solutions, there
exists in $(M_{12})$ at least an element $G_{i'j'}(z) \ne 0$ (note that
$|j'-i'| >1$ since $i'<i<j'<j$). The compatibility between the shapes
(\ref{matshape}) of the matrix $G(z)$ for the pairs of indices $(i,j)$ and
$(i',j')$ implies that (1) $G_{jj} = 0$, (2) $j'-i' = j-i$ and (3) the
elements $G_{i'+1,j'+1}$, \ldots, $G_{i-1,j-1}$ are non-zero. More
generally, one can prove by recursion that if $G_{ij}$ and $G_{ji} \ne 0$,
the non-zero off-diagonal elements of the matrix $G(z)$ lie on two lines
parallel to the diagonal of the matrix and containing the elements $G_{ij}$
and $G_{ji}$ respectively. The inspection of the different configurations
shows that the matrix $G(z)$ is never invertible. 

\medskip\noindent
$\bullet$ \textit{The case (ii)} (i.e. $|j-i|>1$, with $i = 1$ or $j = \enne$
exclusively) requires special attention. A careful examination of the
shape (\ref{matshape}) of the matrix $G(z)$ shows three possibilities:
\begin{itemize}
\item[--]
\vskip -9pt
There exists a pair $(i',j')$ of indices for which $G_{i'j'} \ne 0$ such
that $|j'-i'|>1$ and we are back to case (i);
\item[--]
\vskip -9pt
There is no such $(i',j')$ pair, and $\enne>4$. Then,
it is impossible to have exactly one non-zero off diagonal element in
each row and in each column for an invertible matrix $G(z)$;
\item[--]
\vskip -9pt 
For $\enne=4$, an explicit calculation exhibits the solution
(\ref{eq:solsup}).
\end{itemize}
$\bullet$ \textit{We consider now the case (iii)}, where $i=1$ and $j=\enne$. The
$E_{m1} \otimes E_{m\enne}$ component of (\ref{rsrs}) gives for $m \ne
1,\enne$ (using the fact that $G_{\enne 1}(z) = -qz^{-4}G_{1\enne}(z)$)
\begin{equation}
z^2 (q^2z^2-w^2)\,G_{mm}(z)\,G_{1\enne}(w) - w^2 
(q^2w^2-z^2)\,G_{mm}(w)\,G_{1\enne}(z) = 0
\label{eq:Em1EmN}
\end{equation}
while the $E_{m1} \otimes E_{\enne m}$ component reads
\begin{equation}
z^2\,G_{mm}(z)\,G_{1\enne}(w) - w^2\,G_{mm}(w)\,G_{1\enne}(z) = 0
\label{eq:Em1ENm}
\end{equation}
Equations (\ref{eq:Em1EmN})--(\ref{eq:Em1ENm}) imply $G_{mm} = 0$ for all
$m \ne 1,\enne$. If $G_{\enne\enne} \ne 0$, the $E_{m\enne} \otimes
E_{\enne n}$ component for $m,n \ne 1,\enne$ leads to $G_{mn}(z) =
G_{mn}(w)$, which implies $G_{mn}(z) = 0$ when $1<(m,n)<\enne$ due to
(\ref{eq:kmn}). Therefore the matrix $G(z)$ is not invertible. Hence, we
assume that all diagonal elements are zero. If there exists a pair
$(i',j')$ of indices such that $|j'-i'|>1$, we are back to the case (i)
(and the matrix $G(z)$ is not invertible in that case). Otherwise, all
non-zero off diagonal elements belong to $2 \times 2$ submatrices
corresponding to the indices $(m,m+1)$ of the matrix $G(z)$, denoted
hereafter as `block $m$'. The blocks should not overlap (i.e. a block $m$
cannot be followed by a block $m+1$) and should be adjacent (i.e. a block
$m$ should be followed by a block $m+2$), otherwise the matrix $G(z)$ is
not invertible. Since $G_{1\enne}(z) \ne 0$, the matrix $G(z)$ contains
blocks $m$ with $m$ even only. Using lemma \ref{lemma:DKD}, the matrix
$G(z)$ is brought to the form (\ref{solSNP:AG}). Finally, all remaining
equations obtained from (\ref{rsrs}) are satisfied.

\medskip\noindent
$\bullet$ \textit{We finally consider the case (iv)}, where $j=i+1$. Suppose that
$G_{mm} \ne 0$ for some $m \ne i,j$. Since $j-i=1$, one has necessarily
$m<i,j$ or $m>i,j$. The $E_{im} \otimes E_{mj}$ component of (\ref{rsrs})
reduces here to
\begin{equation}
\big( G_{ij}(z) - G_{ij}(w) \big) \, G_{mm} = 0
\label{eq:EimEmjdz}
\end{equation}
which implies that $G_{ij}(z)$ is constant. 
Moreover, the projection on $E_{mi} \otimes
E_{mj}$ takes here the form
\begin{equation}
\big( z^2w^2\,\delta_{m<i} + \delta_{m>i} \big) G_{ij} \, G_{mm} = 0 
\label{eq:EmiEmjdz}
\end{equation}
leading to $G_{ij} \, G_{mm} = 0$, which is obviously not satisfied.
Therefore, one has $G_{mm} = 0$ for $m \ne i,j$. If $G_{jj} \ne 0$,
$G_{ij}(z)$ is given by (\ref{eq:Kij}) with $\beta_{ij} = 0$. The
projection on $E_{jn} \otimes E_{mj}$ (with $m,n \ne j,j-1$) reads
\begin{equation}
\Big( \big( z^2\,\delta_{j<m} + w^2\,\delta_{j>m} \big) G_{mn}(z) - \big(
z^2\,\delta_{j<n} + w^2\,\delta_{j>n} \big) G_{mn}(w) \Big) G_{jj} = 0
\label{eq:EjnEmjdz}
\end{equation}
and thus leads to $G_{mn}(z) = \mu'_{mn}$ if $m,n<i$ or $m,n>j$, $G_{mn}(z)
= \mu'_{mn}z^2$ if $m<i$ and $n>j$, and $G_{mn}(z) = \mu'_{mn}z^{-2}$ if
$m>j$ and $n<i$. This is clearly in contradiction with (\ref{eq:kmn}),
unless $G_{mn}(z) = 0$ for all $m,n \ne i,j$, giving a non-invertible
matrix.

Therefore, all diagonal elements must be equal to zero. There are then
three possibilities: either there exists a pair $(i',j')$ of indices such
that $|j'-i'|>1$ corresponding to the case (i) or (ii); or $G_{1\enne}(z)
\ne 0$ and we are back to the case (iii); or $G_{1\enne}(z) = 0$ and all
non-zero off diagonal elements consist in blocks $m$ subjected to the same
constraints as in case (iii), but now with $m$ odd. One finds then the
solution (\ref{solSNP:G}) if $\enne$ is even, while the matrix $G(z)$ is
not invertible if $\enne$ is odd.
\finproof

The solutions (\ref{solSNP:z}) were given for the first time in \cite{gand}, and 
 used in \cite{dema} in the context of sine-Gordon and affine Toda
 field theories on the half line. 
The solution (\ref{solSNP:G}) was used in \cite{moras} to define and
study the quantum twisted Yangians. To our 
knowledge, the solutions (\ref{solSNP:AG}), 
(\ref{eq:solsup}) and (\ref{eq:solsupdiag}) are new.
Let us stress that (\ref{eq:solsupdiag}) is the only anti-diagonal matrix
appearing in the classification. Hence, it corresponds, after the
change (\ref{eq:ktilde-k}), to the only case where the reflection equation
(\ref{newTwY}) admits a diagonal solution.
Note that with the use of
 non-invertible diagonal matrices, the lemma \ref{lemma:DKD} provides also
 classes of non-invertible solutions. 

Of course as in the case of SP boundary conditions the ultimate goal is the
derivation of the spectrum of the transfer matrix. However, in our case due
to the lack of diagonal solutions of the equation (\ref{newTwY}), deriving the
spectrum is turning to an intricate problem. This is primarily due to the
fact that an obvious reference state is not available anymore. One could
proceed using a generalization of the method presented in \cite{nepo,chinese}
for the case of non-diagonal boundaries, such analysis however will not be
attempted in the present work. It is under investigation.

\section{Conclusion}

This paper is concerned with generalisations of the XXZ model. We deal
with spin chains where at each site may be associated a different
representation of $\cU_q(gl(\enne))$, and which have various types of
boundary conditions: periodic, soliton preserving or soliton non-preserving.
We have computed the Bethe equations, which is the main physical
results of this paper, thanks to the analytical Bethe ansatz. This
method involves the classification of the representations and the
knowledge of the center of the underlying algebra. Therefore, an
important part of this paper is devoted to the study of the algebra
$\cU_q(\wh{gl}(\enne))$ and its associated subalgebras: the reflection
algebra and the quantum twisted Yangian. For the paper to be
self-contained, we first review these well-known algebraic structures.
Then, we also establish  new results (concerning representations and
centers) for these algebras.
These results look interesting by
themselves from the algebraic point
of view.

Although we treat a generic case, the Bethe equations take
surprisingly a very compact form, depending only on the Drinfel'd
polynomials, the boundaries and the Cartan matrix of $gl(\enne)$.
This simple form enlightens the deep connection between the theory of
 representations and the Bethe ans\"atze. The determination of
 transfer matrix
eigenvectors, which is beyond the scope of analytical Bethe ansatz,
would certainly be simplified by an utter understanding of this
relation.

As mentioned in the paper, we do not deal with non-diagonal boundary
conditions: they cannot be handled, at the moment, with the method
presented here. 
This problem has been treated in some particular cases:  the XXZ
model (i.e. $\cU_q(\wh{gl}(2))$ with fundamental representations)
has been solved in \cite{nepo,chinese}, and some progress has been
achieved for $\cU_q(\wh{gl}(\enne))$ with fundamental representations 
in \cite{YangZhang}. However,
a generic method (similar to the one used for spin chains based on
Yangians \cite{byebye}), which would deal with the full type of
models, is still lacking.

Another open problem is worth mentioning: there exists no systematic
method to compute local Hamiltonians from the transfer matrices for
generic representations. It is known that, in the fundamental
representation, one
obtains the Hamiltonian for the closed or open XXZ model with a very simple
formula. In the general framework treated in this paper, a local
Hamiltonian must be found case by case, after having chosen a
representation at each site. It implies in general the fusion
procedure for auxiliary spaces, and no general proof for the
existence of such local Hamiltonian is known when the representations at
each site are different. This point is crucial, since locality is an important
property in physics. The proof and the explicit construction of local 
Hamiltonian is thus a challenging open question.  
The coefficients of the transfer matrices introduced here being
non-local Hamiltonians, one could be disappointed by such a result.
However, the models being integrable and finite dimensional,
it is reasonable to believe
that the local Hamiltonian is `hidden' in some complicated way in the series of
Hamiltonians provided by the transfer matrix. In particular, it is
known that the Bethe ansatz equations presented in this paper are the 
right ones which will parametrize the spectrum of this hypothetical 
local Hamiltonian.
\\
\\
\textbf{Acknowledgements:} This work is supported by the TMR Network
`EUCLID. Integrable models and applications: from strings to condensed
matter', contract number HPRN-CT-2002-00325.

\appendix

\section{$R$ matrices\label{sect:Rmatrix}}
\subsection{The $R$ matrix of $\cU_{q}({gl}(\enne))$\label{sect:Rfini}}
The $R$ matrix of the finite dimensional quantum group $\cU_{q}({gl}(\enne))$
is given by
\begin{eqnarray}
  && R = q\sum_{a=1}^{{\cal N}} E_{aa}
  \otimes E_{aa}+ \sum_{1\leq a\neq b \leq{\cal N}}
  E_{aa} \otimes E_{bb} +(q-q^{-1})
    \sum_{1\leq a<b\leq{\cal N}} E_{ab}\otimes E_{ba} \,,
\label{Ruqfin}
\end{eqnarray}
where $E_{ab}$ are the elementary matrices with 1 in position $(a,b)$ and 0
elsewhere.

This $R$ matrix obeys the Yang-Baxter equation
\begin{equation}
  R_{12}\ R_{13}\ R_{23} =R_{23}\ R_{13}\ R_{12}.
\end{equation}
If we note $R_{12}[q]$ the matrix (\ref{Ruqfin}) with deformation
parameter $q$, one has the properties
\begin{eqnarray}
    R_{12}[q]\,R_{12}[q^{-1}]=\II\otimes\II\,,\\
    R_{12}[q]-R_{21}[q^{-1}]=(q-q^{-1})\cP\,, \label{eq:R12-R21inv}
\end{eqnarray}
where $\cP\equiv\cP_{12}=\cP_{21}$ is the permutation operator
    \begin{equation}
  \label{eq:P12}
  \cP = \sum_{a,b=1}^\enne E_{ab} \otimes E_{ba} \;,
\end{equation}
and
    \begin{equation}
R_{21} =\cP\,R_{12}\,\cP \,.
\end{equation}

It is also convenient to introduce a deformation of the permutation
operator, the $q$-permutation operator $P_{12}^q\in
End(\CC^\enne\otimes\CC^\enne)$ defined by
\begin{equation}
    P_{12}^q= \sum_{i=1}^\enne E_{ii}\otimes E_{ii} +q \sum_{i> j}^\enne
  E_{ij}\otimes E_{ji}+q^{-1} \sum_{i<j}^\enne E_{ij}\otimes E_{ji}\,.
\label{qpermut}
\end{equation}
We have gathered below some useful properties of the $q$-permutation.
Then, the $R$ matrix (\ref{Ruqfin}) can be rewritten as
\begin{equation}
    R_{12}=\II\otimes\II+q\cP-P_{12}^{q}\,. \label{Ruqfin2}
 \end{equation}
Finally, let us note the following identities
\begin{eqnarray}
R_{21} &=& R_{12}^{t_{1}t_{2}} =
U_{1}U_{2}\,R_{12}\,U^{-1}_{1}U^{-1}_{2}
\label{truc}\\
U_{1}\,R_{12}^{t_{2}}\,U_{1}^{-1} &=& U^t_{2}\,R_{12}^{t_{1}}\,
(U^t_{2})^{-1} \\
\left(R_{12}^{-1}\right)^{t_{1}} &=& M_{1}^{-1}\,\left(R_{12}^{t_{1}}\right)^{-1}
\, M_{1} \label{Rinv-t-inv}
\end{eqnarray}
where
$^{t_{1}}$ (resp. $^{t_{2}}$) is the transposition in the
first (resp. second) space, $U$ is
 any antidiagonal invertible matrix $U=\sum_{a=1}^N u_{a}E_{a\bar a}$
 and
 \begin{equation}
M=\sum_{a=1}^{\enne}\theta_{0}\,q^{\enne-2a+1}\,E_{aa}\,,\quad
\theta_{0}=\pm1\,. \label{def:M}
\end{equation}
Relation (\ref{truc}) implies
\begin{equation}
W_{1}\,W_{2}\,R_{12}\ =\ R_{12}\,W_{1}\,W_{2}
\end{equation}
where $W$ is any invertible diagonal matrix.

\subsection{Some properties of the $q$-permutation\label{sect:qperm}}

 The $q$-permutation can be rewritten in a condensed way as
 \begin{equation}P^q_{12}=\sum_{a,b=1}^\enne q^{sgn(a-b)}E_{ab}\otimes E_{ba}\end{equation}
 where $sgn$ is the sign function, with the convention $sgn(0)=0$.
 Note that this $q$-permutation is not symmetric anymore,
\begin{equation}
 P^q_{21}=\cP\,P^q_{12}\,\cP=P_{12}^{\bar q}\qmbox{with} \bar q=q^{-1}
 \end{equation}
 but still obey the permutation group relations:
 \begin{eqnarray}
 &&\left(P_{12}^q\right)^2=\II\otimes\II\,, \label{Pqcarre}\\
 &&P_{12}^q\,P_{23}^q\,P_{12}^q = P_{23}^q\,P_{12}^q\,P_{23}^q\,.
 \label{eq:permGroup}
 \end{eqnarray}
 More generally, for $q$ and $t$  deformation
 parameters, one has:
 \begin{equation}
 P_{12}^q\,P_{12}^t=P_{12}^{{q}/{t}}\,\cP=\cP\,P_{12}^{t/q}\,.
 \end{equation}
With $P_{12}^q$ comes the notion of \textit{$q$-deformed
operator}: to each
 operator-valued matrix
 \begin{eqnarray}
     A=\sum_{i,j=1}^\enne A_{ij}E_{ij}\,,
 \end{eqnarray}
 one associates its $q$-deformed version as
 \begin{eqnarray}
     A^q=\sum_{i,j=1}^\enne q^{sgn(j-i)}A_{ij}E_{ij}\,.
     \label{def:q-op}
 \end{eqnarray}
 These definitions are justified by the following relations (with
 $\bar q=q^{-1}$, $A_{1}=A\otimes\II$, $A_{2}=\II\otimes A$):
\begin{eqnarray}
  P_{12}^q\,A_{1}\,P_{12}^q=A_{2}^q \qmbox{and}
 P_{12}^q\,A_{2}\,P_{12}^q=A_{1}^{\bar q}\,.
\end{eqnarray}
We will also need:
\begin{eqnarray}
&&  tr_{2}A_{1}\,B_{2}\,P_{12}^q = A\,B^{\bar q}\qmbox{;}
tr_{1}A_{2}\,B_{1}\,P_{12}^q = A\,B^{q}\\
&&  tr_{2}P_{12}^q\,A_{2}\,B_{1} = A^{\bar q}\,B\qmbox{;}
  tr_{1}P_{12}^q \,A_{1}\,B_{2}= A^q\,B\qquad
\end{eqnarray}
as well as, when $[A_{1}\,,\,B_{2}]=0$:
\begin{eqnarray}
  tr_{1}A_{1}\,B_{2}\,P_{12}^q = B\,A^q\qmbox{;}
  tr_{2}A_{2}\,B_{1}\,P_{12}^q = B\,A^{\bar q}\\
  tr_{1}P_{12}^q\,A_{2}\,B_{1} = B^{q}\,A\ \qmbox{;}
  tr_{2}P_{12}^q \,A_{1}\,B_{2}= B^{{\bar q}}\,A\,.
\end{eqnarray}

\subsection{The $R$ matrix of $\cU_{q}(\wh{gl}(\enne))$
\label{app:R-UqAff}}

We consider the $\cU_{q}(\wh{gl}(\enne))$ algebra, whose $R$ matrix can
be constructed through the Baxterization of the $R$ matrix
(\ref{Ruqfin}):
\begin{equation}
R_{12}(z)=z\,R_{12}-z^{-1}\,R_{21}^{-1}
\end{equation}
It is given by the explicit expression
\begin{eqnarray}
 R(z) = {\mathfrak a}(z) \sum_{a=1}^{{\cal N}} E_{aa}
  \otimes E_{aa}+ {\mathfrak b}(z) \sum_{1\leq a\neq b \leq{\cal N}}
  E_{aa} \otimes E_{bb} +{\mathfrak c} \sum_{1\leq a\neq b \leq{\cal N}} z^{
  sgn(b-a)} E_{ab}\otimes E_{ba} \label{r}
\end{eqnarray}
with ${\mathfrak a}(z)=qz-q^{-1}z^{-1}$, ${\mathfrak b}(z)
= z-z^{-1}$ and ${\mathfrak c} = q-q^{-1}=\fa(1)$. It can be also written
as
\begin{equation}
R_{12}(z)=\fb(z)\,\Big(\II\otimes\II-P^q_{12}\Big)+\fa(z)\,\cP\,.
\label{r-joli}
\end{equation}

At $z =1$ the $R$ matrix reduces to the permutation operator $\cP$
given in (\ref{eq:P12}): $R(1)={\mathfrak c}\,\cP$.

In what follows we shall make use of the antidiagonal matrix
$V=\sum_{a,b=1}^\enne V_{ab}E_{ab}$
defined by:
\begin{equation}
  V_{ab} = q^{\frac{a-b}{2}} \delta_{b \bar a} \qmbox{or}
  V_{ab} = i (-1)^{a} q^{\frac{a-b}{2}} \delta_{b \bar a}~.
\end{equation}
where $\bar a = {\cal N}+1 -a $. The second choice is forbidden for
${\cal N}$ odd. We will parametrize these two choices by a sign
$\theta_{0}=\pm1$, the second choice being associated to
$\theta_{0}=-1$. With this convention, one can encompass (formally)
the two choices as
\begin{equation}
  V_{ab} = \theta_0^{a+\frac{1}{2}}q^{\frac{a-\bar a}{2}}\, \delta_{b \bar a}
    \equiv \theta_{a}\,\delta_{b \bar a}.
 \end{equation}
Note that we have in both cases $\theta_{0}^\enne=1$ and $V^2=\II$.

The $R$ matrix (\ref{r}) satisfies the following properties: \\[2mm]

\textit{Yang--Baxter equation} \cite{mac,yang,baxter,korepin}
\begin{equation}
  R_{12}(\frac{z}{w})\ R_{13}(z)\ R_{23}(w)
  =R_{23}(w)\ R_{13}(z)\  R_{12}(\frac{z}{w})
  \label{YBE}
\end{equation}

\begin{equation}
R_{21}(z) =\cP\,R_{12}(z) \,\cP =
R_{12}^{t_{1}t_{2}}(z)
\end{equation}

\textit{Unitarity}
\begin{eqnarray}
  R_{12}(z)\ R_{21}(z^{-1}) = \zeta(z) \label{uni1}
\end{eqnarray}
where
\begin{equation}
\zeta(z) =
(qz-q^{-1}z^{-1})(qz^{-1}-q^{-1}z) \label{def:zeta}
\end{equation}

\textit{Crossing-unitarity}
\begin{eqnarray}
  R_{12}^{t_{1}}(z)\ M_{1}\ R_{12}^{t_{2}}(z^{-1}\rho^{-2})\
  M_{1}^{-1} = \bar\zeta(z\rho)
  \label{croun}
\end{eqnarray}
where  $\rho =q^\frac{\enne}{2}$, $M =V^{t}\,V$
and
\begin{eqnarray}
\bar\zeta(z) =(z\rho-z^{-1}\rho^{-1})(z^{-1}\rho-z\rho^{-1})\,.
\label{def:zetabar}
\end{eqnarray}

\textit{CP-invariance}
\begin{eqnarray}
  R_{21}(z)= U_{1}U_{2}\,R_{12}(z)\,U^{-1}_{1}U^{-1}_{2}
\label{CP-inv}
\end{eqnarray}
for any antidiagonal invertible matrix $U=\sum_{a=1}^N u_{a}E_{a\bar a}$.
Examples of such
matrices are given by $V$, $V^t$ and also $\sum_{a=1}^N E_{a\bar a}$.
\\
This relation implies in particular
\begin{equation}
  \Big [ W_{1} W_{2},\ R_{12}(z) \Big ] =0\,,\label{CP-inv2}
\end{equation}
for any invertible diagonal constant matrix $W$. It can be checked by 
direct calculation that this property remains valid when $W$ is not
invertible.
An example for an (invertible) $W$ matrix is provided by $M=V^{t}\,V$,
 whose explicit expression is given by relation (\ref{def:M}).

The $R$ matrix can be interpreted physically as a scattering matrix
\cite{zamo, korepin, faddeev} describing the interaction between two
solitons that carry the fundamental representation of $gl(\enne)$.

\subsection{The matrix $\bar R(z)$\label{sect:Rbar}}

The CP-invariance of $R$ allows the existence, in the general case, of anti-solitons
carrying the conjugate representation of $gl(\enne)$. The
 scattering matrix which describes the interaction between a
soliton and an anti-soliton, is given by

\begin{eqnarray}
  R_{\bar 12}(z) = \bar R_{12}(z) &=& V_{1}
  R_{12}^{t_{2}}(z^{-1}\rho^{-1})V_{1}
\label{cross}
\\
&=& V_{2}^{t} R_{12}^{t_{1}}(z^{-1}\rho^{-1})  V_{2}^{t} \;=:\; R_{1
\bar 2}(z) = \bar R_{21}^{t_{1}t_{2}}(z). \label{eq:br}
\end{eqnarray}
Note that equality between the first and second lines of these
relations is a consequence of the properties listed above.
In the case $\enne=2$ and for the
second choice of $V$ ($sp(2)$ case), $\bar R$ is proportional to $R$, so
that there is no genuine notion of anti-soliton. This reflects the fact
that the fundamental representation of $sp(2)=sl(2)$ is self-conjugate.
This does not contradict the fact that for $\enne=2$ and for $\theta_0=+1$
($so(2)$ case), there exists a notion of soliton and anti-soliton.

The equality between $R_{\bar 12}(z)$ and $R_{1 \bar 2}(z)$ in
(\ref{cross}) reflects the CP invariance of $R$, from which one also
has $R_{\bar 1 \bar 2} = R_{12}$, i.e. the scattering matrix of two
anti-solitons is equal to the scattering matrix of two solitons. The
explicit form of the $\bar R$ matrix  is
\begin{equation}
  \bar R(z) = \bar {\mathfrak a}(z) \sum_{a=1}^{{\cal N}}
  E_{\bar a \bar a} \otimes E_{aa}+ \bar {\mathfrak b}(z)
  \sum_{1\leq a\neq b \leq{\cal N}} E_{\bar a \bar a} \otimes E_{bb} + {\mathfrak c}
  \sum_{1\leq a\neq b \leq{\cal N}} (\theta_{0}q)^{a-b}
  (z\rho)^{sgn(b-a)}
  E_{\bar b \bar a} \otimes E_{ba}
\label{br}
\end{equation}
where  we set $\bar {\mathfrak a}(z) ={\mathfrak a}(z^{-1}\rho^{-1})$ and
$\bar {\mathfrak b}(z) ={\mathfrak b}(z^{-1}\rho^{-1})$.

The matrix $\bar R(z)$ reduces to a one dimensional projector at $z
=\rho^{-1}$, i.e.
\begin{equation}
  \bar R_{12}(\rho^{-1}) = (q-q^{-1}) \ V_{1}\ {\cal P}_{12}^{t_{2}}\ V_{1} =
  (q-q^{-1})\ {\cal N}\ Q_{12}
  \label{projQ}
\end{equation}
The matrix $Q_{12}$ is a projector (i.e. $Q^{2} =Q$) onto a
one-dimensional space and is written as
\begin{equation}
  Q_{12}= \frac{1}{{\cal N}}\sum_{a, b =1}^{{\cal N}} (\theta_0q)^{a-b}
    E_{\bar b \bar a}\otimes E_{ba}
\end{equation}
Let us remark that $Q_{12}$ is not a symmetric operator i.e.
$Q_{12}\neq Q_{21}$. It will be important in the quantum contraction
for the reflection algebra (see section \ref{sect:fQ}).

The $\bar R$ matrix (\ref{eq:br}) also obeys
\\[2mm]
\textit{(i) A Yang--Baxter equation}
\begin{equation}
  \bar R_{12}(\frac{z}{w})\ \bar R_{13}(z)\
  R_{23}(w) =R_{23}(w)\ \bar R_{13}(z)\ \bar
  R_{12}(\frac{z}{w})
  \label{YBE2}
\end{equation}
\textit{(ii) Unitarity}
\begin{equation}
  \bar R_{12}(z)\ \bar R_{2 1}(z^{-1}) = \bar\zeta(z)
  \label{uni2}
\end{equation}
\textit{(iii) Crossing-unitarity}
\begin{equation}
  \bar R_{12}^{t_{1}}(z)\ M_{1}\ \bar
  R_{12}^{t_{2}}(z^{-1}\rho^{-2})\ M_{1}^{-1}=\zeta(z\rho) \;
  \label{croun2}
\end{equation}
and moreover
\begin{equation}
  \Big [ M_{1} M_{2},\ \bar R_{12}(z) \Big ] =0.
\end{equation}
We remind that the functions $\zeta$ and $\bar\zeta$ are defined in
(\ref{def:zeta}) and (\ref{def:zetabar}).

\section{Review on fusion \label{sect:fQ}}

We present in this appendix the fusion procedure for both periodic
and open spin chains. Such process provides sets of constraints
facilitating the derivation of the spectrum of the corresponding
spin chain. The presentation will be different to the previous one
made for example in \cite{doikou1, doikou2}. Here, we focus on
certain algebraic aspects which may be usefull in other contexts
since we give an explicit construction of central elements for
$\cU_{q}(\wh{gl}(\enne))$ and the reflection algebra.

\subsection{Quantum contraction for $\cU_{q}(\wh{gl}(\enne))$}

The fusion relies essentially on the fact that $\bar R(\rho^{-1})$
reduces to an one-dimensional projector (see (\ref{projQ})). For
this purpose, let us introduce, for $\cL^\pm(z)$ obeying the relations
(\ref{rtt})-(\ref{rtt2}) at $c=0$,
\begin{equation}
  \wh{\cL}^\pm_1(z) = V_{1}\
  \Big(\big({\cL}^\pm_1(z\rho)\big)^{-1}\Big)^{t_{1}}\ V_{1}
\end{equation}
which satisfy
\begin{eqnarray}
 \bar R_{12}(\frac{z}{w})\ \wh{\cL}^\pm_{1}(z)\
  {\cL}^\pm_{2}(w) = {\cL}_{2}^\pm(w)\
  \wh{\cL}^\pm_{1}(z)\ \bar R_{12}(\frac{z}{w})
  \label{funda2}\\
 \bar R_{12}(\frac{z}{w})\ \wh{\cL}^-_{1}(z)\
  {\cL}^+_{2}(w) = {\cL}_{2}^+(w)\
  \wh{\cL}^-_{1}(z)\ \bar R_{12}(\frac{z}{w}).
 \end{eqnarray}
By considering $\frac{z}{w} =\rho^{-1}$ in (\ref{funda2}), we
conclude that
\begin{eqnarray}
\label{eq:qcontL}
  Q_{12} \wh{\cL}^\pm_{1}(z)\ {\cL}^\pm_{2}(z\rho)\
 &=&{\cL}^\pm_{2}(z\rho)\wh{\cL}^\pm_{1}(z)\ Q_{12}
 =Q_{12} \wh{\cL}^\pm_{1}(z)\ {\cL}^\pm_{2}(z\rho)Q_{12}\\
&\equiv&Q_{12}~ \delta(\cL^\pm(z)) \label{deltaT}
\end{eqnarray}
The coefficients of the formal series $\delta(\cL^\pm(z))$, called
quantum contraction of $\cL^\pm(z)$, belong to the center of
$\cU_{q}(\wh{gl}(\enne))$, the proof being similar to the one used in
the case of the Yangian \cite{MNO}.

Using the above construction of central elements for
$\cT^\pm=\Delta^{(\enne)}(\cL^\pm)$ and taking the trace over the
auxiliary spaces $1$ and $2$ of the  relation
\begin{equation}
 \wh\cT^\pm_{1}(z)\cT^\pm_{2}(z\rho)
 = Q_{12}\ \wh{\cT}^\pm_{1}(z)\cT^\pm_{2}(z\rho)+ (
 \II_{\enne}\otimes\II_{\enne}-Q_{12})
 \wh{\cT}^\pm_{1}(z)\cT^\pm_{2}(z\rho)\;,
 \label{fusbe}
\end{equation}
we obtain the fusion relation
\begin{equation}
\label{fusionQ} \wh{t}^\pm(z)t^\pm(z\rho) =
\delta(\cT^\pm(z))+\wt{t}^\pm(z)\;.
\end{equation}
We have introduced the fused transfer matrix
\begin{equation}
  \wt{t}^\pm(z) = \mbox{tr}_{12}\Big((
 \II_{\enne}\otimes\II_{\enne}-Q_{12})
 \wh{\cT}^\pm_{1}(z)\cT^\pm_{2}(z\rho)
 \Big) \label{fusedtm}
\end{equation}
and the transfer matrix
\begin{equation}
\wh{t}^\pm(z)=\tr_1(\wh{\cT}^\pm_{1}(z))\;. \label{tr11}
\end{equation}

\subsection{Quantum contraction for the reflection algebra}

As in the closed case we are going to exploit the fact that the $R$
matrix reduces to a one-dimensional projector for a particular
value of the spectral parameter. We need also new solutions, denoted
$\bar K(z)$, of the same reflection equation (\ref{re}) describing
physically the reflection of an anti-soliton to an anti-soliton (see
also \cite{doikou2,selene}). A new type of reflection equation is
needed,
which gives an additional constraint between $K(z)$ and $\bar
K(z)$
\begin{equation}
  \bar R_{12}(\frac{z}{w})\ \bar K_{1}(z)\ \bar
  R_{21}(zw)\ K_{2}(w)=
  K_{2}(w)\ \bar R_{12}(zw)\ \bar
  K_{1}(z)\ \bar R_{21}(\frac{z}{w}),
  \label{re2}
\end{equation}
Note that this equation is equivalent (exchanging spaces 1 and 2,
arguments $z$ and $w$, and using unitarity for $\bar
R(z)$) to
\begin{equation}
  \bar R_{12}(\frac{z}{w})\ K_{1}(z)\ \bar
  R_{21}(zw)\ \bar K_{2}(w)=
  \bar K_{2}(w)\ \bar R_{12}(zw)\ 
  K_{1}(z)\ \bar R_{21}(\frac{z}{w}),
\end{equation}
which shows that couple of solutions $(K(z),\bar K(z))$ can be
consistently defined. \\
\begin{lemma}
Let $K(z)$ and $\bar K(z)$ be two diagonal solutions (\ref{eq:Kdiag}) of the
reflection equation (\ref{re}), with parameters $(\cM,\xi)$ and
$(\bar\cM,\bar\xi)$ respectively. Then, they obey the additional relation
(\ref{re2}) if  and only if
\begin{equation}
\cM+\bar\cM=\cN \qquad\mbox{and}\qquad 
\xi\,q^{\cM/2}=\pm\,\bar\xi\,q^{\bar\cM/2}
\end{equation}
\end{lemma}
\underline{Proof:} 
Up to a multiplication by $(zw)^{2}$, the relation (\ref{re2}) is a
polynomial in $z$ and $w$. Considering the $(zw)^{2}$ coefficient,
one gets
\begin{eqnarray}
&&\bar\xi^{2}\left(\sum_{a<b}q^{2b-\enne-1}\,(k_{b}-k_{a})\ \cE_{ab}
+(1-q^{-2})\sum_{a<b<c}q^{2c-\enne-1}\,(k_{b}-k_{a})\ \cE_{ab}\right)
\nonu
&&\qquad\qquad=
(\rho\xi)^{2}\left(\sum_{a<b}(\wt k_{a}-q^{-2}\wt k_{b})\ \cE_{ab}
+(1-q^{-2})\sum_{a<c<b}\wt k_{b}\ \cE_{ab}\right)
\end{eqnarray}
where $\cE_{ab}=E_{ab}\otimes E_{ab}-E_{ba}\otimes E_{ba}$ and 
$k_{a}$ (resp. $\wt k_{a}$) are the diagonal terms of $K(0)$ (resp.
 $V\bar K(0)V^t$). \\
Considering the case $b=a+1$, we are led to
\begin{equation}
q^{-1}\bar\xi^{2}(k_{a+1}-k_{a}) = \xi^{2}(\wt k_{a}-q^{-2}\wt k_{a+1}) 
\label{eq:KKbar}
\end{equation}
Since $k_{a+1}-k_{a}=0$ if $a\neq\cM$ and $\wt k_{a}-q^{-2}\wt
k_{a+1}=0$ if $a\neq \cN-\bar\cM$, this implies the first constraint. The
second one is obtained setting $a=\cM$ in (\ref{eq:KKbar}).\\
Finally, one checks directly that these constraints are sufficient to solve the
relation (\ref{re2}).\finproof
In particular, $K(z)=\II$ and $\bar K(z)=\II$ are
solutions to the set of constraints.\\
Similarly, we introduce a new solution $\bar K^+(z)$ of the dual
reflection equation (\ref{red}) with the following consistency
relation
\begin{eqnarray}
  && \bar R_{12}(\frac{w}{z})\ \Big(\bar K_{1}^{+}(z)\Big)^t\
  M_{1}^{-1}\ \bar R_{21}(w^{-1}z^{-1}\rho^{-2})\ M_{1}\
  \Big(K_{2}^{+}(w)\Big)^t= \nonumber \\
  && \Big(K^{+}_{2}(w)\Big)^t\ M_{1}\ \bar
  R_{12}(w^{-1}z^{-1}\rho^{-2})\ M_{1}^{-1}\
  \Big(\bar K_{1}^{+}(z)\Big)^t\ \bar R_{21}(\frac{w}{z})\;.
 \label{red2}
\end{eqnarray}
Starting from $K(z)$ and $\bar K(z)$, solutions to equations
(\ref{re}), (\ref{red}) and (\ref{re2}), the matrices
$K^+(z)=f(z)K(\rho^{-1}z^{-1})^tM$ and $\bar K^+(z)=g(z)\bar
K(\rho^{-1}z^{-1})^tM$ are solutions to the dual reflections
equations ($f(z)$ and $g(z)$ are arbitrary functions chosen in the
spin chain context such that the matrices have analytical entries).
Similarly to relation (\ref{KgenN}), we can define a new monodromy
matrix
\begin{equation}
\bar \cB_1(z)= \wh{\cT}^+_1(z)~\bar K_1(z)~
V_1^{t}\Big(\cT^-_1(z^{-1}\rho^{-1})\Big)^{t_1}V_1^{t}\;,
\label{barKgenN}
\end{equation}
obeying the exchange relation (\ref{re}) and satisfying consistency
relation (\ref{re2}) with $\cB(z)$. We can deduce from relation
(\ref{re2}) (multiplying on the right by $V_1V_2$ and setting
$\frac{w}{z}=\rho$) the following equality
\begin{eqnarray}
\label{eq:qcontB} Q_{12}\bar \cB_{1}(z)\ \bar
  R_{21}(z^2\rho)\ \cB_{2}(z\rho)V_1V_2&=&
  \cB_{2}(z\rho)\ \bar R_{12}(z^2\rho)\ \bar
  \cB_{1}(z)\ V_1V_2\ Q_{12}\hspace{2cm}\\
  &\equiv& Q_{12}\delta(\cB(z))
  \label{deltab}
\end{eqnarray}
and from relation (\ref{red2}) (transposing in both spaces 1
and 2, multiplying on the left by $V_1V_2\bar R_{21}(\frac{z}{w})$
and on the right by $\bar R_{12}(\frac{z}{w})$ and setting
$\frac{z}{w}=\rho$)
\begin{eqnarray}
&&\hspace{-1cm}Q_{12}V_1V_2  K^+_{2}(z\rho)\ M_1\bar
  R_{12}(z^{-2}\rho^{-3})\ M_1^{-1} \bar K^+_{1}(z)\\
  &&\hspace{1cm}=
  V_1V_2 \bar K^+_{1}(z)\ M_1^{-1}
  \bar R_{21}(z^{-2}\rho^{-3})\ M_1
  K^+_{2}(z\rho)\ \ Q_{12}
  \equiv Q_{12}\delta(K^+(z))
\end{eqnarray}
\begin{proposition}
All the coefficients of the series $\delta(\cB(z))$, called quantum
contraction of $\cB(z)$, belong to the center of the reflection
algebra.
\end{proposition}
\underline{Proof:} Using reflection equations (\ref{re}) and
(\ref{re2}) as well as Yang-Baxter equations (\ref{YBE}) and
(\ref{YBE2}), we can show
\begin{eqnarray}
&&\hspace{-1cm}\bar
R_{01}(\frac{w}{z})R_{02}(\frac{w}{y})\cB_0(w)R_{20}(wy) \bar
R_{10}(wz)\cB_2(y)\bar R_{12}(yz)\bar \cB_1(z)\bar
R_{21}(\frac{z}{y})=\nonumber\\
&&\hspace{1cm}\cB_2(y)\bar R_{12}(yz)\bar \cB_1(z)\bar
R_{21}(\frac{z}{y})\bar R_{01}(wz)
R_{02}(wy)\cB_0(w)R_{20}(\frac{w}{y}) \bar R_{10}(\frac{w}{z})
\end{eqnarray}
Multiplying on the right by $V_1V_2$ and taking the particular value
$y=\rho z$ in the previous relation, we obtain
\begin{eqnarray}
&&\hspace{-1cm}\bar R_{01}(\frac{w}{z})R_{02}(\frac{w}{\rho
z})\cB_0(w)R_{20}(\rho wz) \bar
R_{10}(wz)Q_{12}\delta(\cB(z))=\nonumber
\\
&&\hspace{3cm}\delta(\cB(z))Q_{12}V_1V_2\bar R_{01}(wz) R_{02}(\rho
wz)\cB_0(w)R_{20}(\frac{w}{\rho z}) \bar R_{10}(\frac{w}{z})V_1V_2
\label{vald}
\end{eqnarray}
By a direct computation, one can show the
following properties
\begin{eqnarray}
Q_{12}V_1V_2\bar R_{01}(wz) R_{02}(\rho wz)V_1V_2
=\mathfrak a(\rho wz) \mathfrak a(\frac{1}{\rho zw})Q_{12}
=R_{20}(\rho wz)\bar R_{10}(wz)Q_{12}
\label{vald1}\\
Q_{12}V_1V_2R_{20}(\frac{w}{\rho z}) \bar R_{10}(\frac{w}{z})V_1V_2
=\mathfrak b(\frac{\rho w}{z}) \mathfrak b(\frac{z}{\rho w})Q_{12}
=\bar R_{01}(\frac{w}{z})R_{02}(\frac{w}{\rho z})Q_{12}
\;.\label{vald2}
\end{eqnarray}
Using these relations, equality (\ref{vald}) can be rewritten as
\begin{eqnarray}
\cB_0(w)\delta(\cB(z))Q_{12}=\delta(\cB(z))\cB_0(w)Q_{12}\;,
\end{eqnarray}
which is equivalent to the statement of the proposition.
\finproof

 {From} the particular form (\ref{KgenN}) of $\cB(z)$ and
(\ref{barKgenN}) of $\bar \cB(z)$, we deduce that
\begin{equation}
\delta(\cB(z))=\delta\big(\cT^+(z)\big)~ \delta\big(K(z)\big)~
\delta\big(\cT^{-}(z^{-1})^{-1}\big)
\end{equation}
where $\delta(\cT^+(z))$, $\delta\big(\cT^{-}(z^{-1})^{-1}\big)$ are
defined by (\ref{deltaT}) and
$\delta\big(K(z)\big)$ by (\ref{deltab}).\\
Now, we can obtain the fused relation, using similar relation to
(\ref{fusbe}),
\begin{equation}
\zeta(z^2\rho^2)\bar{b}(z)b(z\rho)= \delta(K^+(z))\,\delta(\cB(z)) +
\wt{b}(z)
\end{equation}
where
\begin{eqnarray}
\bar{b}(z)&=&\tr_1\Big(\bar K_1^+(z)\bar
\cB_1(z)\Big) \label{def:bbar}\\
 \wt{b}(z)&=&\tr_{12}\Big((\II_{\enne}\otimes\II_{\enne}-Q_{12})V_1V_2  K^+_{2}(z\rho)\ M_1\bar
  R_{12}(z^{-2}\rho^{-3})\ M_1^{-1} \bar K^+_{1}(z)\nonumber\\
  &&\hspace{3cm}\times\bar \cB_{1}(z)\ \bar
  R_{21}(z^2\rho)\ \cB_{2}(z\rho)V_1V_2\Big)
\end{eqnarray}
The transfer matrix $\wt{b}(z)$ is the so-called fused transfer
matrix.

\section{Generalised fusion \label{sect:gefu}}

We describe a generalised fusion procedure for $\cU_{q}(gl({\enne}))$ open
and closed spin chains \cite{lepetit}. The procedure we use follows the
lines of the construction of the quantum determinant for
$\cU_{q}(gl({\enne}))$, the Sklyanin determinant for quantum twisted Yangians
\cite{moras} and reflection algebras. The crucial observation here is that
for the general case a one-dimensional projector can be obtained by
repeating the fusion procedure ${\enne}$ times (recall that ${\enne}^{
\otimes {\enne}} = 1\oplus\ldots$). The procedure described in the previous
section is basically a consequence of the fact that ${\enne} \otimes \bar
{\enne} = 1\ \oplus \ ({\enne}^2-1)$. Let us now introduce the following
necessary objects for the \emph{generalised} fusion procedure for the
closed and open spin chains (see also equations (2.13) and (2.14) in
\cite{avdo}).

\subsection{$q$-antisymmetriser}

We start with the $q$-permutation (\ref{qpermut}): it
 provides an action of the symmetric group $\mathfrak{S}_\enne$ on the
space $(\CC^\enne)^{\otimes \enne}$ in the following way. To each
transposition $(i,i+1)\in \mathfrak{S}_\enne$, we associate the
$q$-permutation $P^q_{i}$ $q$-permuting $i^{th}$ and $(i+1)^{th}$
spaces. Then, we set, for $\sigma\in\mathfrak{S}_\enne$,
$P^q_\sigma=P^q_{i_1}\dots P^q_{i_l}$ where
$\sigma=(i_1,i_1+1)\dots(i_l,i_l+1)$ is a reduced decomposition. The
decomposition does not depend on the choice of the reduced
decomposition because of the braid relation satisfied by $P^q_i$
(see equations (\ref{Pqcarre}) and {\ref{eq:permGroup})). We recall
that $l=l(\sigma)$ is called length of the permutation $\sigma$. We
can now introduce the $q$-antisymmetriser
\begin{equation}
  \cA^q=\frac{1}{\enne !}\sum_{\sigma\in \mathfrak{S}_\enne}
  (-1)^{l(\sigma)}P^q_{\sigma}
\end{equation}
This operator is the one-dimensional projector on the $q$-antisymmetric
representation belonging to the tensorial product of $\enne$ fundamental
representations of $\cU_{q}(gl({\enne}))$.

The fundamental result to obtain the generalized fusion consists in
writing the $q$-antisymmetriser in terms of a product of R-matrices
\begin{equation}
\label{AR} \cA^q~\prod_{1\leq a<b\leq
\enne}\left(q^{a-b}-q^{b-a}\right) =
\frac{1}{\enne!}~ \prod_{1\leq a<b\leq \enne}R_{ab}(q^{a-b})
\end{equation}
where the product in the right hand side is taken in the
lexicographical order on the pairs $(a,b)$.

\subsection{Fusion from the quantum determinant}

We  define, for $a_1,\dots a_\enne\in\{1,\dots,\enne\}$,
\begin{eqnarray}
  \cL^\pm_{<a_1\dots a_\enne>}= \cL^\pm_{a_1}(z_{a_1}) \ldots
  \cL^\pm_{a_\enne}(z_{a_{\enne}}), \qmbox{with}
    z_{a} =zq^{a-1}\,,\ a=1,\ldots , \enne
  \label{ftt}
\end{eqnarray}
where $\cL^\pm(z)$ are solution of relations (\ref{rtt}) and
(\ref{rtt2}) with $k=0$. Let us remark that we will make the fusion
for a tensor product of auxiliary spaces (here denoted by
$1,\dots,\enne$). It is important to note the difference between
these $\enne$ auxiliary spaces and the $\enne$ quantum spaces
present, for example, in relation (\ref{mono}) or (\ref{KN}).

 {From} relation (\ref{AR}), we can show
\begin{eqnarray}
\cA^q\; \cL^\pm_{<1\dots\enne>}&=&\cL^\pm_{<\enne\dots
1>}\;\cA^q=\cA^q\;
\cL^\pm_{<1\dots\enne>}\;\cA^q\\
 &\equiv&\cA^q\; qdet\,\cL^\pm(z)\;.
 \label{def:qdet}
\end{eqnarray}
Relation (\ref{def:qdet}) defines formal series in terms of $z$
whose the coefficients belong to the center of
$\cU_{q}(\wh{gl}({\enne}))$ (we can also show that they are
algebraically independent and generate the center). Similarly, we
can write
\begin{eqnarray}
qdet\,\cL^\pm(z)&=& \sum_{\sigma\in \mathfrak{S}_N}
(-q)^{-l(\sigma)}L^\pm_{\sigma(1),1}(z)\dots
L^\pm_{\sigma(\enne),\enne}(zq^{\enne-1})
\label{qdet1}\\
&=& \sum_{\sigma\in \mathfrak{S}_N}
(-q)^{l(\sigma)}L^\pm_{1,\sigma(1)}(zq^{\enne-1})\dots
L^\pm_{\enne,\sigma(\enne)}(z)\label{qdet2}
\end{eqnarray}

Using this above construction of the quantum determinant for
$\cT^\pm=\Delta^{(\enne)}(\cL^\pm)$ and taking the trace over all the
auxiliary spaces $1,\dots,\enne$ of the following relation
\begin{equation}
 \cT^\pm_{<1\dots\enne>} = {\cA}^q\ \cT^\pm_{<1\dots\enne>}+
 (\id -\cA^q)
 \cT^\pm_{<1\dots\enne>}\;,
\end{equation}
we obtain the so-called generalized fused relation
\begin{equation}
\label{fusion2} \prod_{l=1}^{{\cal N}} t^\pm(z_{l}) =
qdet\,\cT^\pm(z)+\wt{t}^\pm(z)\;.
\end{equation}
We have defined the so-called fused transfer matrix
$\wt{t}^\pm(z)=\tr_{<1\dots\enne>}
(\id -\cA^q)\cT^\pm_{<1\dots\enne>}$.
To prove this relation, one needs to use
$\tr_{<1\dots\enne>}\cA^q=1$. The equation (\ref{fusion2})
plays a crucial role in determining
the spectrum of the periodic $\cU_{q}(gl({\cal N}))$ spin chain.

\subsection{Fusion from the Sklyanin determinant}

We can use a similar construction for the reflection algebra to
obtain its central elements and a fusion relation. The
 quantum determinant is replaced by the so-called Sklyanin determinant,
defined as follows
\begin{eqnarray}
\cA^q\; \cB_{<1\dots\enne>}&=&\cB_{<\enne\dots 1>} \;\cA^{q}=\cA^q\;
\cB_{<1\dots\enne>}\;\cA^{q}\\
 &\equiv&\cA^q\; sdet\,\cB(z)\;.
 \label{def:sdet}
\end{eqnarray}
where
\begin{eqnarray}
\cB_{<a_1\dots a_\enne>}=\prod_{k=1,...,\enne}^{\longrightarrow}
\left(
  {\cB}_{a_k} (z_{a_k}) \left(\prod_{l=k+1,...,\enne}^{\longrightarrow}
  R_{a_l a_k}(z_{a_l}z_{a_k}) \right) V_{a_{k}}^t\right)\;.
  \label{sol1t}
\end{eqnarray}
In the above relations, $z_a=zq^{a-1}$ and ${\cB}(z)$ is the
generating function for the quantum reflection algebra defined by the
 relation (\ref{re}). Relation (\ref{def:sdet}) defines
formal series in terms of $z$ whose the coefficients belong to the
center of $\cR$. To study the fusion, it is necessary to also
introduce the quantity
\begin{eqnarray}
{K}^{+}_{<a_1\dots a_\enne>} &\!=\!&
  \prod_{k=1,\dots,\enne}^{\longleftarrow} \left(V_{a_{k}} \left(
  \prod_{l=k+1,\dots,\enne}^{\longleftarrow} R_{a_k
  a_{l}}(z_{a_k}^{-1}z_{{a_l}}^{-1}\rho^{-2})\right) M_{a_k}^{-1}
  K^+_{a_k}(z_{a_k})\right)
  \label{frr2}
\end{eqnarray}
where $K^+(z)$ is solution of relation (\ref{red}) and we can show
that
\begin{eqnarray}
\cA^q\; K^+_{<1\dots\enne>}&=&
K^+_{<\enne\dots 1>}\;\cA^q=\cA^q\; K^+_{<1\dots\enne>}\;\cA^q\\
 &\equiv&\cA^q\; sdet\,K^+(z)\;.
 \label{def:sdetK+}
\end{eqnarray}
Taking the trace over all the auxiliary spaces $1,\dots,\enne$ of
the following relation
\begin{equation}
 K^+_{<1\dots\enne>}\cB_{<1\dots\enne>} =
 {\cal A}^q\ K^+_{<1\dots\enne>}\cB_{<1\dots\enne>}+
 (\id -{\cal A}^q)
 K^+_{<1\dots\enne>}\cB_{<1\dots\enne>}\;,
 \label{eq:aIA}
\end{equation}
we obtain the so-called generalized fused relation for the
reflection algebra
\begin{equation}
\label{fusion3}
\left(\prod_{a<b}\bar\zeta(\frac{\rho^{-1}}{z_{a}z_{b}})\right)\,\prod_{l=1}^{{\cal N}} b(z_{l}) =
sdet\,K^+(z)~sdet\,\cB(z)+\wt{b}(z)\;,
\end{equation}
where $\bar\zeta(z)$ is defined in (\ref{def:zetabar}).
We recall that $b(z)=\tr_0(K_{0}^+(z)\cB_0(z))$ and we have defined the
so-called fused transfer matrix
$\wt{b}(z)=\tr_{<1\dots\enne>}(\id -\cA^q)
K^+_{<1\dots\enne>}\cB_{<1\dots\enne>}$. As discussed in
\cite{avdo}, the matrix $K^+_{<1\dots\enne>}$ is necessary so that
the trace of the r.h.s. of (\ref{eq:aIA}) decouples to a product of
${\enne}$ transfer matrices. Note that if we choose $\cB(z)$ of the
form (\ref{KgenN}), then its Sklyanin determinant can be determined
in terms of the quantum determinant:
\begin{equation}
sdet\,\cB(z)=sdet\,K(z)~qdet\,\cT^+(z)
\big(qdet\,\cT^-(z^{-1}q^{-\cN+1})\big)^{-1}\;.
\end{equation}
This factorised form of the Sklyanin determinant allows us to prove
easily that its coefficients belong to the center of $\cR$. It is
also very useful to compute its explicit form in a given
representation (see equation (\ref{eq:sdetB})).

\end{document}